\documentclass[onecolumn,12pt]{emulateapj}
\usepackage{apjfonts}
  \usepackage{verbatim}  
  
  \submitted{Accepted for publication in The Astrophysical Journal}
 
\begin{document}

\newcommand{\com}[1]{{\bf #1}}
\newcommand{\bc}{}

\newcommand{\NS}{N_H} 
\newcommand{\UA}{u_1} 
\newcommand{\UB}{u_2} 
\newcommand{\JPF}{P} 
\newcommand{\CDF}{D_{X_0}} 
\newcommand{\DEP}{\delta} 
\newcommand{\offset}[2]{f^{#1}_{#2}} 
\newcommand{\orbc}[2]{c^{#1}_{#2}} 
\newcommand{\op}{\phi} 
\newcommand{\HSTP}{P_{HST}} 
\newcommand{\MIDT}{\langle t \rangle } 
\newcommand{\PH}{H} 
\newcommand{\Dh}{h} 
\newcommand{\RA}{R_0} 

\bibliographystyle{apj}

\title{Near-infrared transit photometry of the exoplanet HD~149026b}

\author{
Joshua A.~Carter\altaffilmark{1},
Joshua N.~Winn\altaffilmark{1},
Ronald Gilliland\altaffilmark{2},
Matthew J. Holman\altaffilmark{3}
}

\email{carterja@mit.edu}

\altaffiltext{1}{Department of Physics, and Kavli Institute for
  Astrophysics and Space Research, Massachusetts Institute of
  Technology, Cambridge, MA 02139}

\altaffiltext{2}{Space Telescope Science Institute, 3700 San Martin
  Drive, Baltimore, MD 21218}

\altaffiltext{3}{Harvard-Smithsonian Center for Astrophysics, 60
  Garden St., Cambridge, MA 02139}

\begin{abstract}
  The transiting exoplanet HD~149026b is an important case for theories
  of planet formation and planetary structure, because the planet's
  relatively small size has been interpreted as evidence for a highly
  metal-enriched composition. We present observations of 4 transits
  with the Near Infrared Camera and Multi-Object Spectrometer on the
  {\em Hubble Space Telescope}\, within a wavelength range of
  1.1--2.0~$\mu$m. Analysis of the light curve gives the most precise
  estimate yet of the stellar mean density, $\rho_\star =
  0.497^{+0.042}_{-0.057}$ g~cm$^{-3}$. By requiring agreement between
  the observed stellar properties (including $\rho_\star$) and stellar
  evolutionary models, we refine the estimate of the stellar radius:
  $R_\star = 1.541^{+0.046}_{-0.042}$ $R_\sun$. We also find a deeper
  transit than has been measured at optical and mid-infrared
  wavelengths. Taken together, these findings imply a planetary radius
  of $R_p = 0.813^{+0.027}_{-0.025}$ $R_{\rm Jup}$, which is larger
  than earlier estimates. Models of the planetary interior still
  require a metal-enriched composition, although the required degree
  of metal enrichment is reduced. It is also possible that the deeper
  NICMOS transit is caused by wavelength-dependent absorption by
  constituents in the planet's atmosphere, although simple model
  atmospheres do not predict this effect to be strong enough to
  account for the discrepancy. We use the 4 newly-measured transit
  times to compute a refined transit ephemeris.
\end{abstract}


\section{Introduction}

Since its discovery by Sato et al.~(2005), HD~149026b has been one of
the most closely scrutinized planets outside the Solar system. It is a
close-in gas giant, orbiting a G star with a period of only 2.5~d.
Observations of transits (Sato et al.~2005, Charbonneau et al.~2006,
Winn et al.~2008b, Nutzman et al.~2008), in combination with
observations of radial-velocity variations of the parent star (Sato et
al.~2005), have shown that the planet has approximately Saturn's mass
but is considerably denser, despite the intense irradiation from the
parent star that should inflate the planet and lower its
density. There is consensus among theorists that the reason for the
``shrunken radius'' is a highly metal-enriched composition, although
the total metal mass, its distribution within the planet, and the
reason for the enrichment are debated [Sato et al.~(2005), Fortney et
al.~(2006), Ikoma et al.~(2006), Broeg \& Wuchterl~(2007), Burrows et
al. (2007)]. The total metal mass, for example, ranges from
$60~M_\earth$ to $114~M_\earth$ among the possible models. The latter
estimate would represent 80\% of the total mass of the planet.

The planet's outer atmosphere is also of interest, given the possibly
unusual composition and the strong heating from the parent star.
Models by Fortney et al.~(2005) indicated the possibility of a very
hot stratosphere as a result of gaseous TiO and VO opacity. By using
the {\em Spitzer Space Telescope}\, to observe a planetary
occultation, Harrington et al.~(2007) found the planet's 8~$\mu$m
brightness temperature to be much larger than the temperature that one
would expect based on thermal equilibrium with the incident stellar
radiation. This may be the result of the predicted TiO and VO heating,
although the details of whether and where these absorbers actually
condense in the atmosphere are not yet understood (Fortney et
al.~2005, Harrington et al.~2007).

Fundamental to all these discussions are the measurements of the mass
and radius of HD~149026b. These measurements are limited by the
uncertainties in the stellar mass and radius. One way to improve the
situation is to observe transits with greater photometric precision
than has been possible before. As shown by Seager \& Mallen-Ornelas
(2003), with a good light curve and Kepler's third law, one may
determine the stellar mean density. If the mean density is known
precisely enough, it is a key constraint that can be combined with the
other stellar observables (parallax, apparent magnitude, effective
temperature, metallicity, etc.) and stellar-evolutionary models to
determine the stellar mass and radius. This technique has been put
into practice for many other systems [see, e.g., Sozzetti et
al.~(2007), Holman et al.~(2007), Torres et al.~(2008)] but never to
advantage for HD~149026b because of the limited precision of prior
determinations of $\rho_\star$ (Winn et al.~2008b, Nutzman et
al.~2008). Observers must cope with the small transit depth of
2.5~mmag (smaller than any other transiting exoplanet by a factor of
two) and the paucity of suitable comparison stars within the field of
view of most telescopes.

In this paper we present observations of transits of HD~149026b with
the Near Infrared Camera and Multi-Object Spectrometer [NICMOS,
Thompson~(1992)] on board the {\em Hubble Space Telescope}\, ({\em
  HST\,}). We chose this instrument because a high precision in
relative photometry is possible even without using comparison stars
(Gilliland~2006, Swain et al.~2008) and because the reduced stellar
limb-darkening at near-infrared wavelengths is advantageous for the
light-curve analysis (Carter et al.~2008, P\'al et al.~2008). We have
organized this paper as follows.  In \S~\ref{sec:obs} we describe the
observations and data reductions leading to the final photometric time
series. In \S~\ref{sec:nicanalysis} we describe our photometric model
and the results of the NICMOS light-curve analysis. In
\S~\ref{sec:stellar}, we describe how the light-curve results were
incorporated into stellar-evolutionary models to determine the
parameters of the HD~149026 system. In \S~\ref{sec:joint}, the light
analysis is repeated using not only the NICMOS data but also the most
precise light curves that have been obtained at optical and
mid-infrared wavelengths. In \S~\ref{sec:timing}, all the available
transit times are analyzed to produce a refined transit ephemeris and
to search for possible period variations that could be indicative of
additional bodies in the HD~149026 system (Holman \& Murray~2005, Agol
et al.~2005, Ford \& Holman 2007). Finally, in
\S~\ref{sec:conclusions}, we discuss the possible implications of our
observations and analysis.

\section{Observations and Reductions} \label{sec:obs}

We observed HD~149026 on four occasions (``visits'' in {\it HST}\,
parlance) when transits were predicted to occur, on 2007~Dec~22,
2007~Dec~24, 2008~Feb~08, and 2008~Mar~20.  Each visit consisted of
five orbits spanning a transit. Between each pair of orbits is an
observing gap of approximately 45 minutes, when {\em HST\,}'s vision
is blocked by the Earth. The visits were scheduled in such a manner
that the combined data set provides complete phase coverage of the
transit, including redundant coverage of the critical ingress and
egress phases. In particular, visits 1 and 3 covered the ingress
phase, and visit 2 covered both ingress and egress phases. Visit 4
captured the beginning of egress.

We used Camera 3 of the NICMOS detector, a $256\times256$ HgCdTe array
with a field of view of $51.2\farcs \times 51.2\farcs$. We used the
G141 grism filter, which is centered at $1.4\mu$m, spans 0.8~$\mu$m
and is roughly equivalent to $H$ band. An exposure was obtained every
13~s. The camera was operated in ``MULTIACCUM'' mode, wherein five
nondestructive readouts are recorded during a single exposure, and the
first readout is subtracted from the final readout. After accounting
for overheads, the effective integration time was 4~s per exposure.
We deliberately defocused the instrument to give a full-width at
half-maximum (FWHM) of approximately 5 pixels in the cross-dispersion
direction. This was done for two reasons: firstly, when focused,
camera 3 undersamples the point-spread--function (PSF) of point
sources; and secondly, the detector pixels exhibit intra-pixel
sensitivity variations as large as 30\%. Defocusing the images causes
the PSF to be well-sampled and averages over the intra-pixel
sensitivity variations.

Approximately 220 exposures were collected during each {\it HST}\,
orbit. Experience with {\it HST}\, has shown that photometric
stability is relatively poor during the first orbit of a given visit.
Our observations were scheduled under the assumption that the first
orbit from each visit would not be utilized, and indeed we ended up
omitting the first-orbit data from our analysis. At the start of each
visit, we obtained a single non-dispersed image using a narrow filter
centered at $1.66$~$\mu$m in order to establish the pixel position
corresponding to zero dispersion. We then adopted the relation from
the {\it HST}\, Data Handbook for NICMOS\footnote{{\tt
    http://www.stsci.edu/hst/nicmos/documents/handbooks/DataHandbookv7/}},
\begin{eqnarray}
	\lambda(\Delta x) = -0.007992~\Delta x + 1.401
\end{eqnarray}
where $\lambda(\Delta x)$ is the wavelength (measured in $\mu$m), and
$\Delta x$ is the $x$ coordinate (measured in pixels) relative to the
center of the undispersed image.

For completeness, we performed the steps of flat-fielding, background
subtraction, and pixel flagging, as described below; however, it is
noteworthy that these steps in the data reduction made very little
difference in the aperture photometry or in the final
results. Flat-field correction for grism images is not straightforward
and is not done as part of the standard NICMOS pipeline reductions,
because the appropriate flat field depends both upon wavelength and
upon the position of the source in the non-dispersed image. To
accomplish the flat-field correction, we obtained seven flat fields,
each using a narrow bandpass within the G141 bandpass, and fitted the
data for each pixel with a quadratic function of wavelength
$C[\lambda; x,y]$.  We then applied $C[\lambda(\Delta x); x, y]$ as a
multiplicative correction to each of our science images, using the
wavelength-coordinate relation $\lambda(\Delta x)$ that was determined
from the single non-dispersed image. The background level was
estimated in each image based on the counts in a relatively clean
region of the detector (away from the spectral trace) and subtracted
from the entire image.  To identify bad pixels, all images from a
given orbit were used to create a time series of counts specific to
each pixel. Pixels showing an anomalously large variance were
flagged. The list of flagged pixels was appended to the list of hot or
cold pixels that were identified in the standard NICMOS pipeline
reductions, and the values of all of those bad pixels were replaced by
interpolated values of the neighboring good pixels.

Aperture photometry was performed on the first-order spectrum, using a
simple sum of the counts within a rectangular box centered on the
spectral trace. The box had a width of 20 pixels in the
cross-dispersion direction (the $y$ direction), which was four times
the FWHM of the PSF. The box had a length of 120 pixels in the
dispersion direction (the $x$ direction), which was long enough the
capture the entire first-order spectrum.

At this stage the data had been reduced to a single number per image:
the total number of counts in the aperture (the ``flux''). We examined
the resulting time series. As expected, the data collected during the
first orbits of each visit showed flux variations that were both
larger in amplitude and different in their time-dependence than the
variations observed in subsequent orbits. The first-orbit data were
excluded from subsequent analysis. In addition, we excluded the 10
exposures near the beginning and end of each orbit sequence, because
they showed strong flux variations that are probably due to ``Earth
shine.''  After these exclusions, there remained $800$, $820$, $800$,
and $792$ good data points in visits 1, 2, 3, and 4, respectively.

\section{NICMOS Light-Curve Analysis} \label{sec:nicanalysis}

\begin{figure}[htbp] 
   \centering
   \epsscale{1.00}
  \plotone{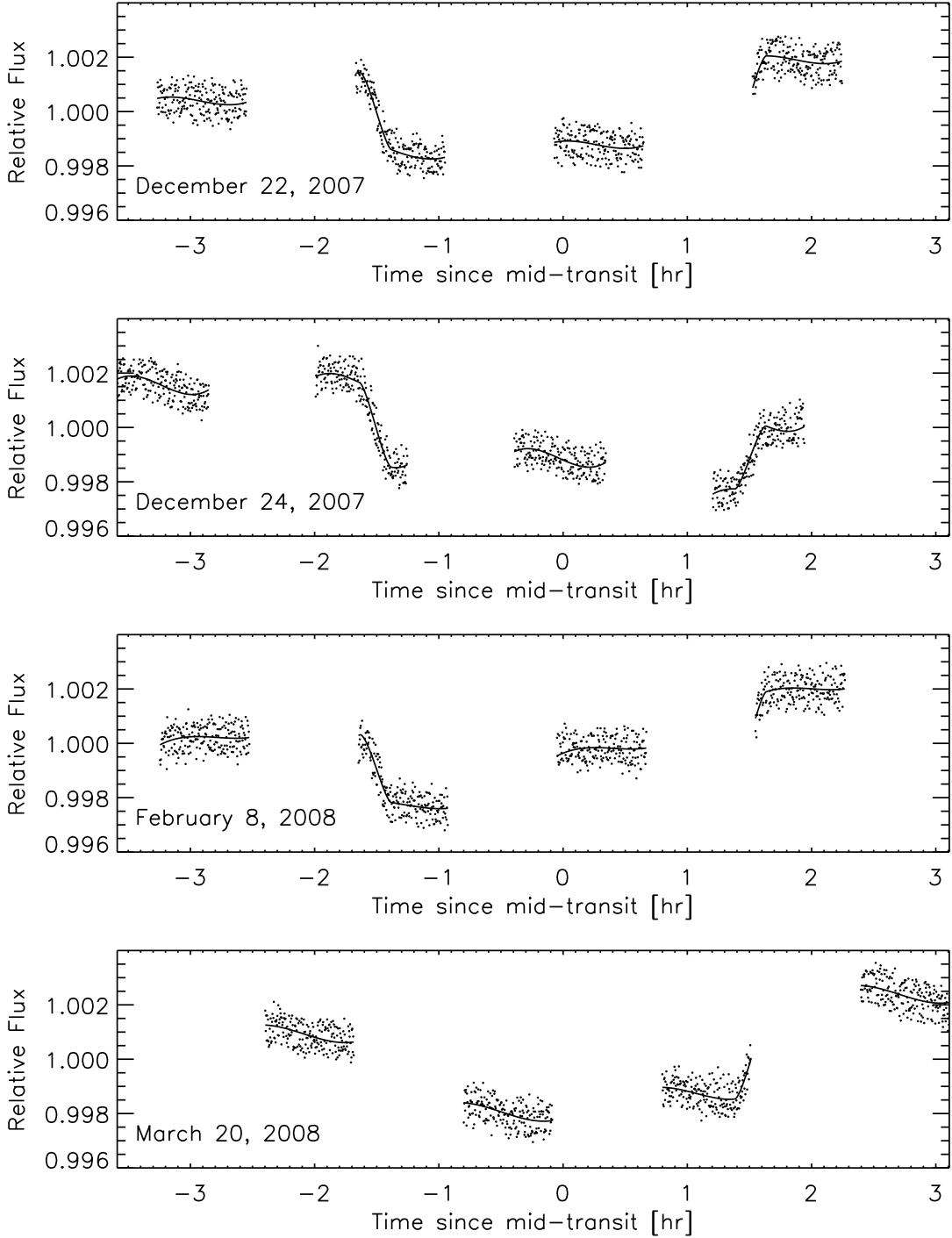}
  \caption{NICMOS photometry (1.1--2.0~$\mu$m) of HD~149026b of 4
    transits, with interruptions due to Earth occultations. Plotted
    are the results of simple aperture photometry. The observed
    variations are a combination of the transit signal and systematic
    effects (intra-orbital variations and inter-orbital
    discontinuities). The solid curve is the best-fitting model that
    accounts for both the transit signal and systematic effects.}
   \label{fig:raw}
\end{figure}

Fig.~\ref{fig:raw} shows the time series of the aperture-summed flux,
after dividing by the mean flux. In each panel, the zero point of the
$x$-axis is the expected mid-transit time. The flux decrement during
the transit is identifiable, but this decrement is superimposed on at
least two other sources of variability: orbit-to-orbit
discontinuities, and smooth intra-orbital variability showing a
consistent pattern among all orbits of a given visit. The
intra-orbital variability has been seen by all other investigators
attempting precise {\it HST}\, photometry of single bright stars,
since the pioneering work by Brown et al.~(2001), and the flux
discontinuities have been seen by other investigators using NICMOS
(see, e.g., Swain et al.~2008). The origins of these systematic
effects have not been established. The orbit-to-orbit consistency of
the smooth variations suggests a phenomenon that is a function of the
phase of the telescope's orbit around the Earth, such as thermal
cycling or scattered light. The discontinuities between orbits suggest
a non-repeating event associated with the re-aquisition of the target
star after each Earth occultation, such as pointing changes or
positional shifts of the grism filter.

Ideally, the underlying physical processes giving rise to these
systematic effects could be ascertained, and this understanding would
lead to either the recognition of an improved method for deriving the
photometric signal or a physical model that could be used to correct
the aperture-summed flux. Given that we do not yet have such
knowledge, what can be done? The intra-orbital variations are very
well-described by a smooth function of the {\it HST}\, orbital phase;
following other investigators we used a smooth function with several
adjustable parameters as an {\it ad hoc}\, model for this variation.
The parameters of this model are fairly well constrained by the
out-of-transit data, for which all variations are assumed to be
systematic effects.

\begin{figure}[htbp] 
   \centering
   \epsscale{1.00}
  \plotone{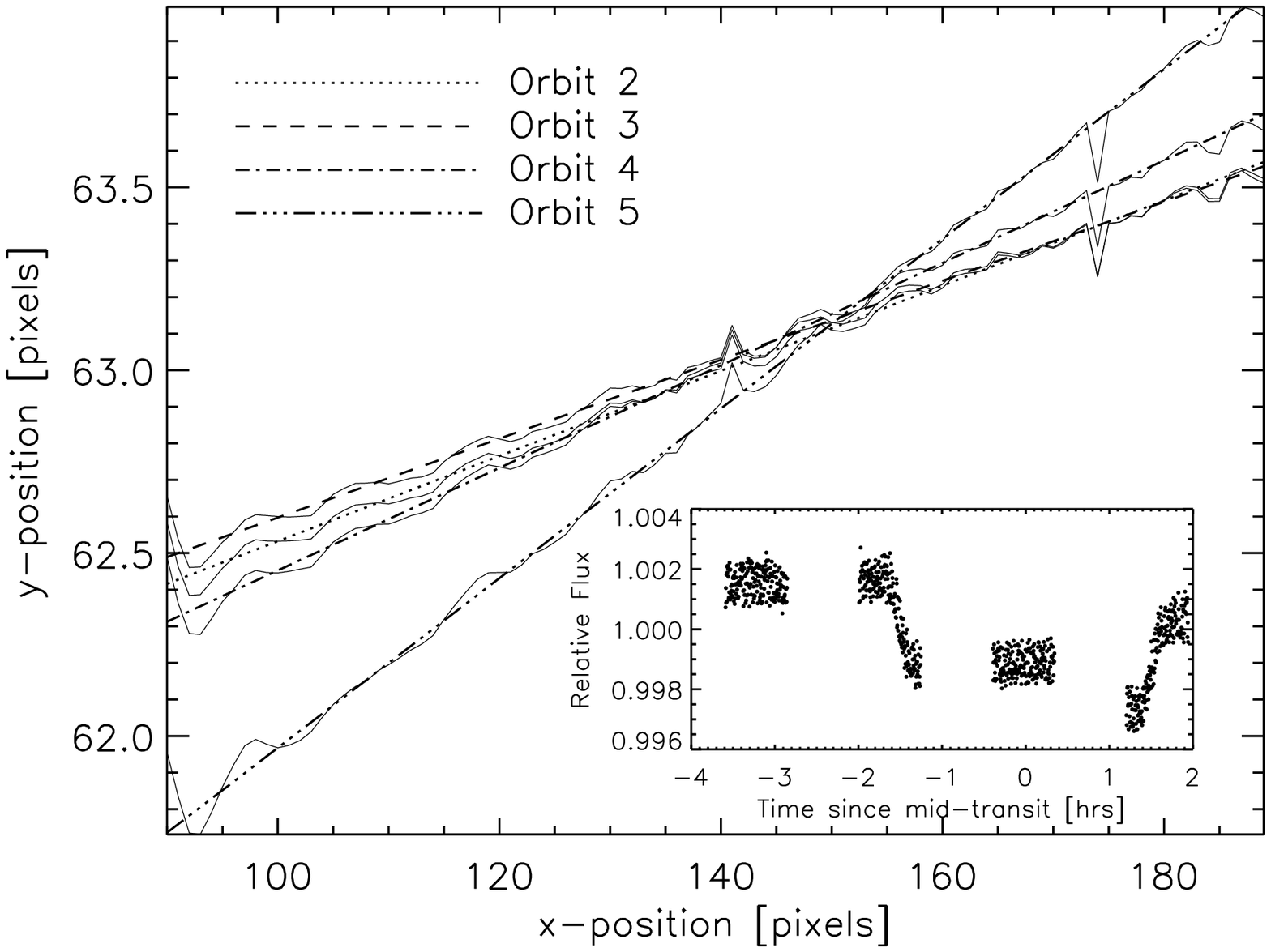}
  \caption{Illustration of inter-orbital variations of the spectral trace.  The solid curves are the flux-weighted mean $y$ position of the first-order spectrum as a function of $x$.  Overplotted are linear fits to $y(x)$.  The inset figure shows the measured light curve after dividing out the intra-orbital variations correlated with {\em HST} orbital phase.  The largest rotation of the spectral trace (at the fifth orbit) coincides with the largest discontinuity in the flux time series.}
  \label{fig:trace}
  \end{figure}

The inter-orbital discontinuities are more problematic. To investigate
the systematic effects, we examined the spectral trace on each
image. Specifically we computed the flux-weighted mean $y$ position as
a function of $x$, giving a curve $y(x)$ representing the centroid of
the spectral trace (the ``footprint'' of the spectrum on the
detector). We also estimated the orientation of the spectral trace
relative to the detector edges, by performing a linear fit to the
previously calculated function $y(x)$. figure~\ref{fig:trace} shows
the results. Within a single orbit, the position and orientation of
the spectral trace are relatively constant, as compared to the larger
movements that are observed between orbits. The largest variations of the
spectral trace (its position and width) seem to coincide with the largest discontinuities in
the flux time series. 

Given the correlations that are observed between the properties of the
spectral trace and the aperture-summed flux, the approach taken by
Swain et al.~(2008) and other investigators is to ``decorrelate'' the
flux against a number of measured parameters (``state variables'')
such as the mean $y$ position, cross-dispersion width, orientation
angle, and so forth. One way to achieve this decorrelation is to fit
linear functions of the state variables to the out-of-transit data,
for which all time variations are expected to be due to the systematic
effects. Then the best-fitting parameters are used to correct all of
the data. Alternatively, one could fit for the linear functions of the
state variables simultaneously with the parameters describing the
transit light curve.

We attempted both of these procedures and found that while they do
reduce the amplitude of the systematic effects, they still leave
highly significant systematic variations. We also find this procedure
to be undesirable because it is not clear which parameters to include
in the fit; because the fitted parameters are highly correlated (the
state variables do not vary independently); and because we have no
justification for the assumption of a linear function for any of these
parameters, without an understanding of the underlying physical
effect. An example of a possibly relevant physical effect that would
not necessarily be described by a linear function is intra-pixel
sensitivity variation, which could lead to a function that is periodic
in the pixel coordinates of the spectral trace.

We attempted to fit numerous physically-motivated models (based on the
premise of intra-pixel sensitivity variations, among others), and did
not find any such model that provided a good fit to the data while
also having only a few, nondegenerate adjustable
parameters. Ultimately we gave up on attempting to correct the
intra-orbital discontinuities based on {\it a priori}\,
information. Instead we included in our model an adjustable
multiplicative factor specific to each orbit. This might seem
devastating to the goal of analyzing the transit light curve, but this
is not so. It means that we cannot make use of the relative flux
between orbits to derive the transit depth; but as we will show, the
truly precious information is in the duration of ingress or egress,
which is much less vulnerable to the problem of flux
discontinuities. In addition, four of the orbits spanned a full
ingress or egress. Data from those four orbits does provide useful
information about the transit depth, because the discontinuities
appear between orbits and not within orbits.

All together, our model for the flux variation due to systematic
effects is
\begin{eqnarray}
  f_{\rm sys}(t) =  \offset{v}{o}
  \left\{1 + \orbc{v}{0}\op(t) + \orbc{v}{1}[\op(t)]^2 + \orbc{v}{2}[\op(t)]^3\right\},
\label{eq:f}
\end{eqnarray}
where the $v$ index specifies the visit number (1--4), the $o$ index
specifies the orbit number (1--4) within each visit after omitting the
first orbit, the 16 numbers $\offset{v}{o}$ are the multiplicative
factors describing the inter-orbital discontinuities, $\op(t)$ is the
{\em HST}\, orbital phase at time $t$, and the 12 numbers $c_i^v$ (3
per visit) are constants specifying a polynomial function of $\phi$
that describes the intra-orbital variation.  The {\em HST}\, orbital
phase was defined as
\begin{eqnarray}
  \op(t) \equiv
   \frac{ \left(t-\MIDT \right) ~{\rm mod}~  \HSTP}{\HSTP},
   \label{eq:phase}
\end{eqnarray}   
where $\HSTP = 1.5975$~hours is the orbital period of the
{\em HST}\, around the Earth and $\MIDT$ is the midpoint of each
orbit's observations. The choice of a polynomial, as opposed to some
other smoothly varying function, was arbitrary. We also tried using
sinusoidal functions with an angular frequency of $2\pi/\HSTP$, with
no significant differences in any of the results described below.
 
For the transit model, we used the analytic formulas of Mandel \&
Agol~(2002). Our parameters were the planet-to-star radius ratio
($R_p/R_\star$), the cosine of the orbital inclination ($\cos i$), the
semimajor axis in units of the stellar radius ($a/R_\star$), the
mid-transit time $t^v_c$, and the two coefficients $\UA$ and $\UB$ of
a quadratic limb-darkening law,
\begin{eqnarray}
	\frac{I_\mu}{I_1} = 1-\UA (1-\mu)-\UB (1-\mu)^2
\end{eqnarray}
where $\mu$ is the cosine of the angle between the observer and the
normal to the stellar surface and $I_\mu$ is the specific intensity as
a function of $\mu$. We allowed $\UA$ and $\UB$ to vary freely,
subject to the conditions $\UA+\UB < 1$, $\UA+\UB > 0$, and $\UA > 0$.
These conditions require the brightness profile to be everywhere
positive and monotonically decreasing from limb to center. In practice,
the fitting parameters were actually
\begin{eqnarray}
	\UA' &=& \UA \cos 40^\circ- \UB \sin 40^\circ \\
	\UB' &=& \UA \sin 40^\circ+\UB \cos 40^\circ
\end{eqnarray}
because $\UA'$ and $\UB'$ are weakly correlated, unlike $\UA$ and
$\UB$ (P\'{a}l~2008). In computing the transit light curve we assumed
the orbit to be circular, consistent with the findings of Sato et
al.~(2005) and Madhusudhan \& Winn (2008). We held the orbital period
fixed at the value $P = 2.87588$ days based on the results of Winn et
al.~(2008b).  Here the period is used only to relate the measured
transit durations and $a/R_\star$. The fractional error in $P$ is
approximately $10^4$ times smaller than the fractional error in
$a/R_\star$ and is safely ignored (Carter et al.~2008).

\begin{figure}[htbp] 
\centering
\epsscale{1.00}
\plotone{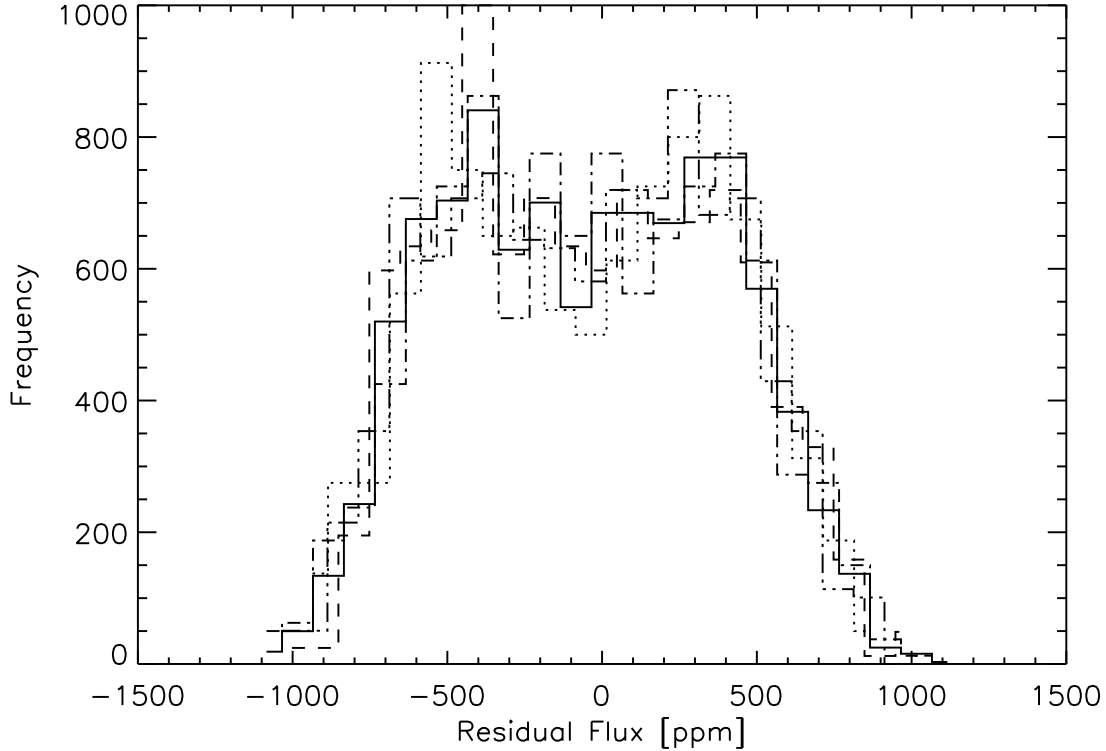}
\caption{Histograms of the residuals between the data and the
  best-fitting model. The solid line is the histogram based on all the
  data. The other lines are for data specific to visit 1 (dotted),
  visit 2 (dashed), visit 3 (dash-dot), and visit 4
  (dash-dot-dot-dot).}
\label{fig:hist}
\end{figure}

Our complete model of the photometric time series was the product of
the transit model and $f_{\rm sys}(t)$ (Eq.~\ref{eq:f}).  We fitted
simultaneously for the parameter set $R_p/R_\star$, $\cos i$,
$a/R_\star$, $\UA'$, $\UB'$, $\{t^v_c\}$, $\{\offset{v}{o}\}$, and
$\{\orbc{v}{i}\})$. The polynomial describing intra-orbital variations
was specific to each visit, and the flux discontinuities were specific
to each orbit, but the transit parameters were required to be
consistent across all orbits and visits. We performed a least-squares
fit to the unbinned data using a box-constrained Levenberg-Marquardt
algorithm (Levenberg~1944, Marquardt~1963, Lourakis~2004) utilizing
the Jacobian calculation of P\'{a}l~(2008). Box constraints were
needed to enforce the restrictions on the limb darkening parameters
$\UA$ and $\UB$. The goodness-of-fit statistic was
\begin{eqnarray}
  \chi^2 & = & \sum_{v=1}^4 \sum_{i=1}^{N_v}
  \left( \frac{f^v_{\rm obs}(t_i)-f^v_{\rm calc}(t_i)}
         {\sigma_v} \right)^2
\label{eq:chi_squared}
\end{eqnarray}
where $f^v_{mod}(i)$ the calculated flux at the time of the $i^{th}$
data point during visit $v$, $f^v_{obs}(i)$ is the $i^{th}$ flux
measurement during visit $v$, $N_v$ is the number of data points in
visit $v$, and $\sigma_v$ was assumed to be a constant at this step.
The solid curve in Fig.~\ref{fig:raw} shows the best-fitting
model. The root-mean-square (rms) residual between the data and the
best-fitting model is 440 parts per million (ppm). This is
approximately $2.2$ times the expected noise level calculated within
the NICMOS calibration pipeline (which is dominated by photon noise). Fig.~\ref{fig:hist} shows histograms of the residuals
for all the data and for each visit individually. The residuals are
not Gaussian; the flattened peak in the histograms indicates a nonzero
kurtosis\footnote{
	One may wonder about the effect of the apparent non-Gaussianity of the
noise, shown in Fig.~3. To investigate this issue we used an Edgeworth
expansion to create a new $\chi^2$-like statistic that accounts for
the skewness and kurtosis of the residuals (see, e.g., Amendola et
al.~1996). Using this different fitting statistic, we found that the
best-fit parameter values were unchanged. This was not surprising
because bias is expected to arise from skewness (as opposed to
kurtosis) and the skewness of the residuals is very small. However,
the confidence intervals are affected by the kurtosis. We found that
accounting for kurtosis leads to error bars that are {\it smaller}\,
than the error bars quoted here, but only by a modest amount
($\lesssim 20\%$). For simplicity, the results quoted in this paper
are based on the standard $\chi^2$ statistic given in Eq.~(7).}.
	
\begin{figure}[htbp] 
\centering
\epsscale{1.00}
\plottwo{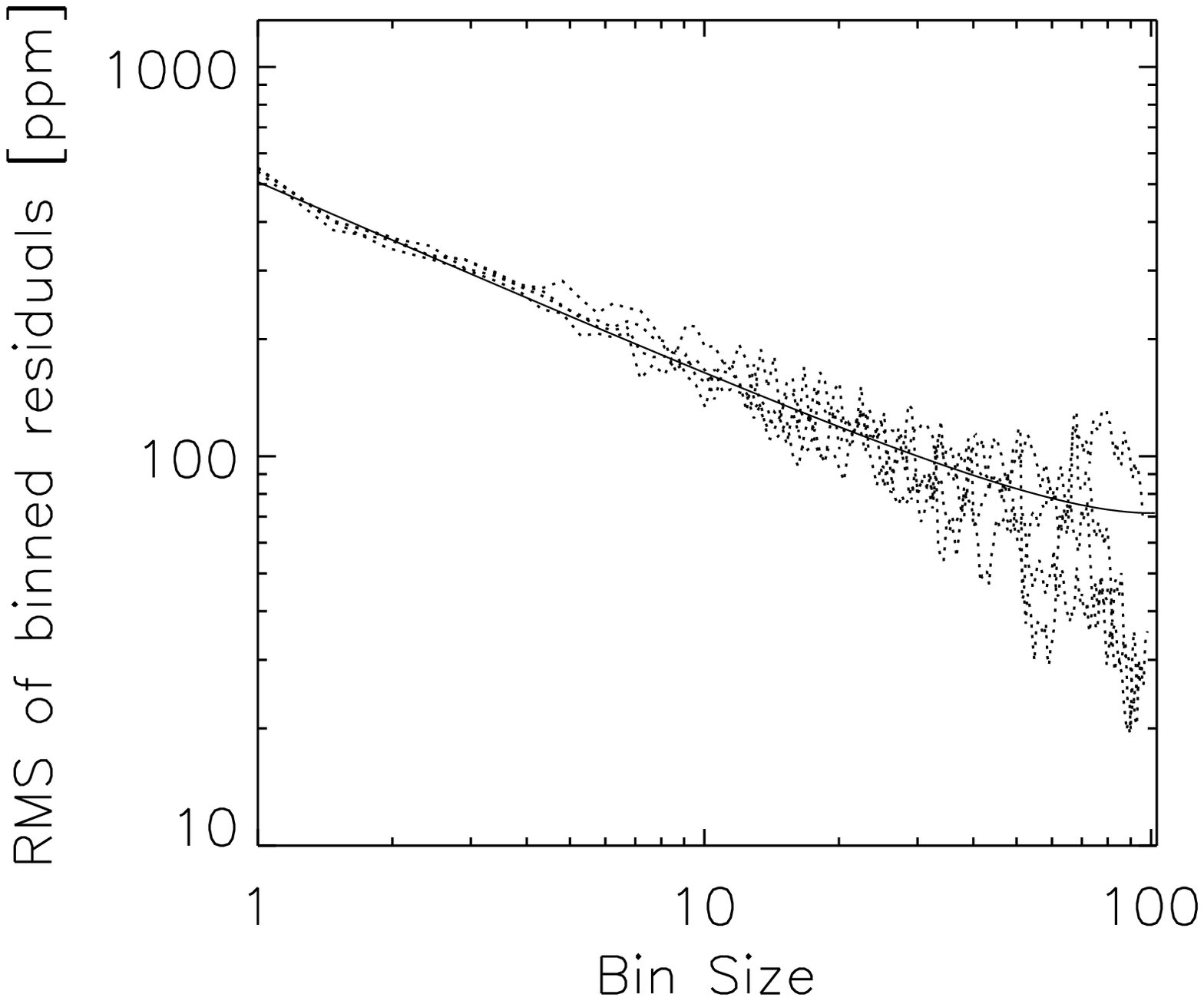}{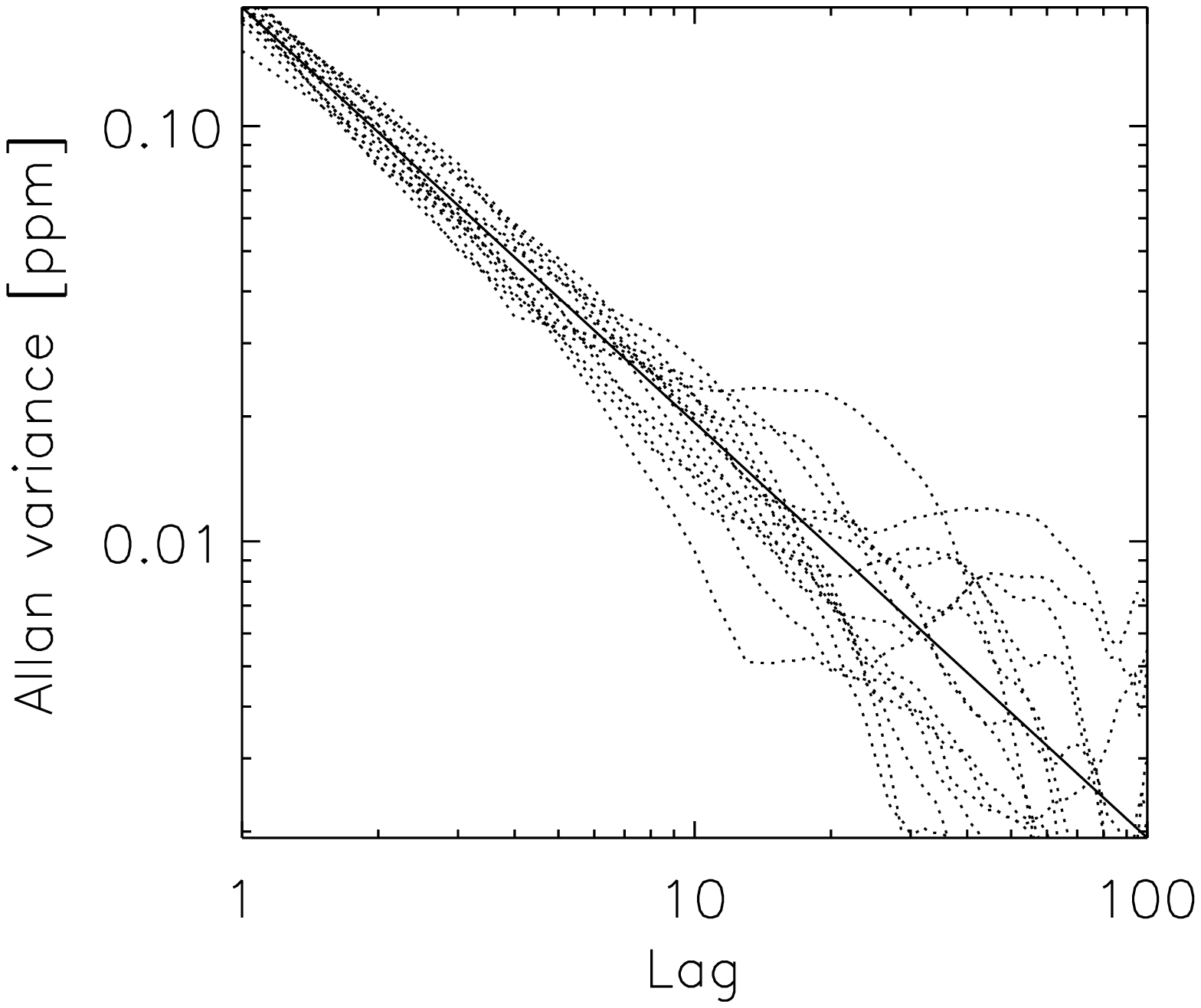}
\caption{Assessment of correlated noise. Left panel: The rms of
  time-binned residuals, as a function of bin size, for each of the 4
  visits. Right panel: The Allan variance of the residuals, as a
  function of lag, for each of the 16 orbits. The dotted lines are the
  results of the calculations based on the data, and the solid lines
  show the expected trend if the noise were uncorrelated.}
\label{fig:rms}
\end{figure}

Evidently the noise is not photon-limited, and is not Gaussian, but at
least it does not appear to be strongly correlated in time. We
assessed the degree of temporal correlations (``red noise'') in two
ways. First, we binned the residuals in time by a factor $N$ ranging
from 1 to 100, and calculated the standard deviation $\sigma_N$ of the
binned residuals. The results are shown in Fig.~\ref{fig:rms}.  They
follow closely the expectation of independent random numbers,
$\sigma_N = \sigma_1~N^{-1/2}~[M/(M-1)]^{1/2}$, where $M$ is the
number of bins. Second, we calculated the Allan (1964) variance
$\sigma_A^2(l)$ of the residuals, defined as
\begin{eqnarray}
  \sigma_A^2(l) =
  \frac{1}{2 (N+1-2l)} \sum_{i=0}^{N-2l} \left( \frac{1}{l} \sum_{j=0}^{l-1} r_{i+j}-r_{i+j+l}\right)^2
\end{eqnarray}
where $r_k$ denotes the residual of the $k$th data point, $N$ is the
number of data points, and $l$ is the lag. The Allan variance is
commonly used in the time metrology literature to assess $1/f$
noise. For independent residuals, one expects $\sigma_A^2(l) \approx
\sigma_A^2(0)/l$. The results for our data, also shown in
Fig.~\ref{fig:rms}, satisfy this expectation. There is no readily
identifiable time-correlated component in the time series of
residuals.

\begin{figure}[htbp] 
   \centering
   \epsscale{1.00}
   \plotone{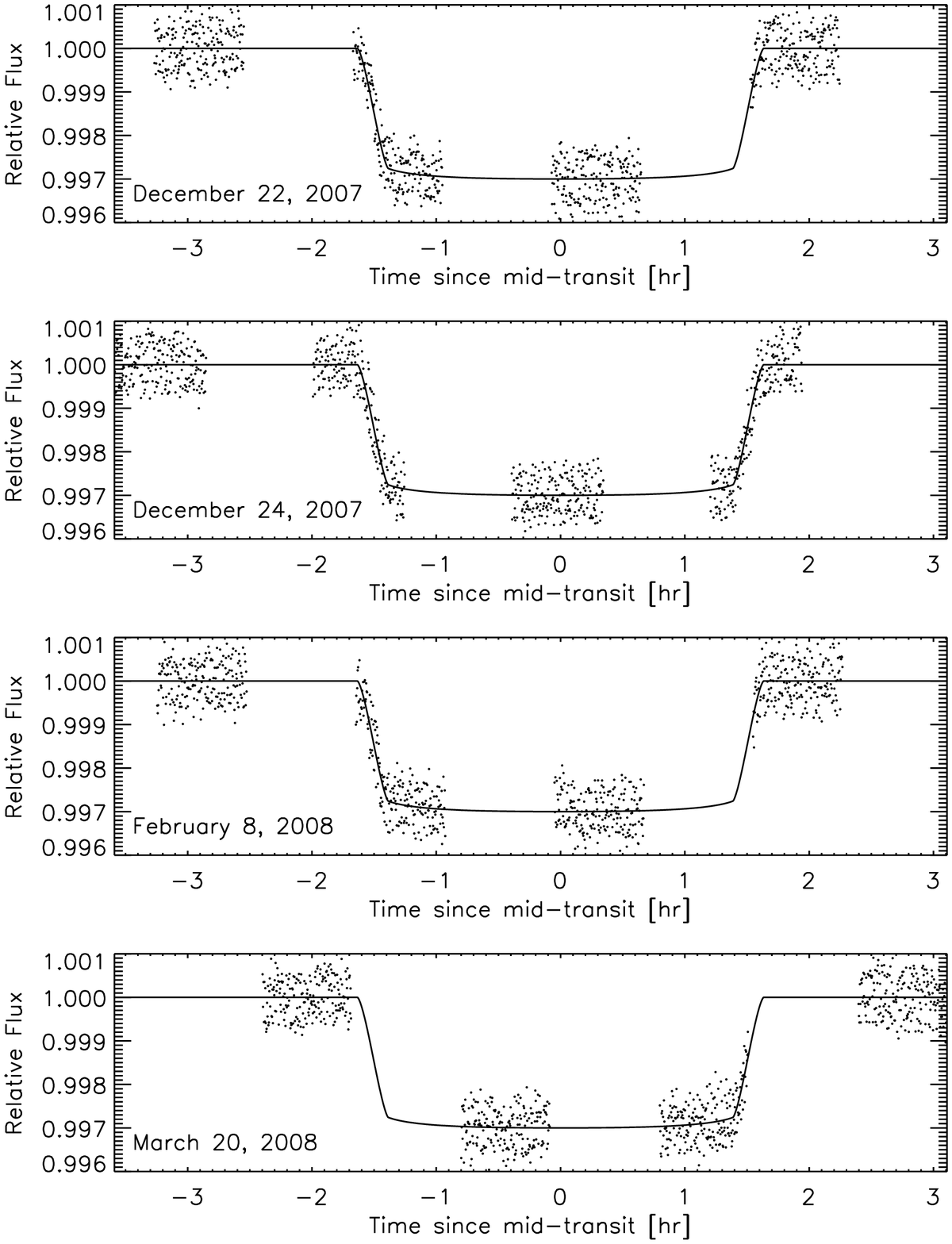}
   \caption{NICMOS photometry (1.1--2.0~$\mu$m) of 4 transits of HD~149026b,
     after correcting for systematic effects. In each
     panel, the solid line shows the best-fitting model.}
   \label{fig:rawperperts}
\end{figure}

\begin{figure}[htbp] 
   \centering
   \epsscale{1.00}
  \plotone{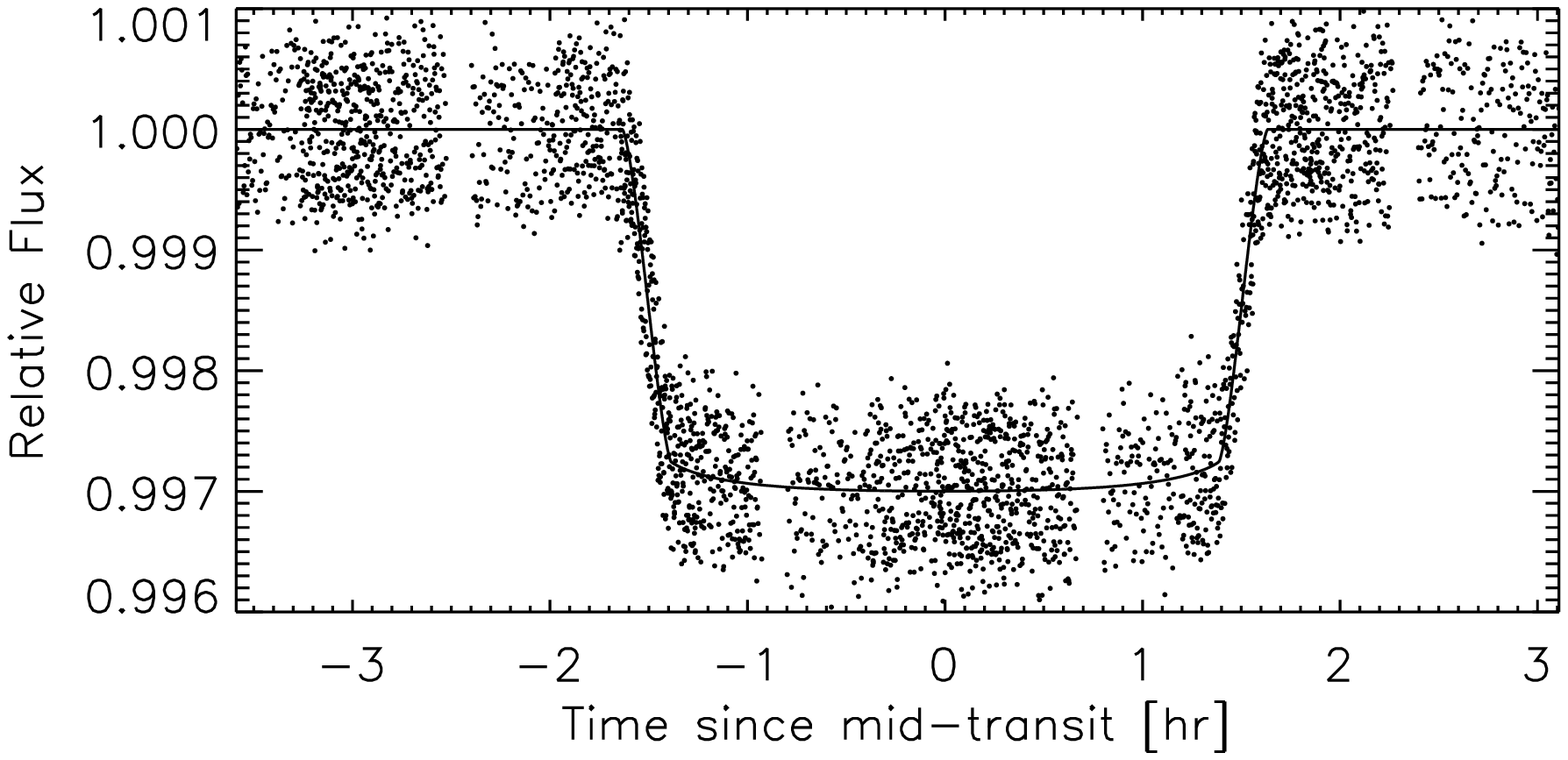}
 \plotone{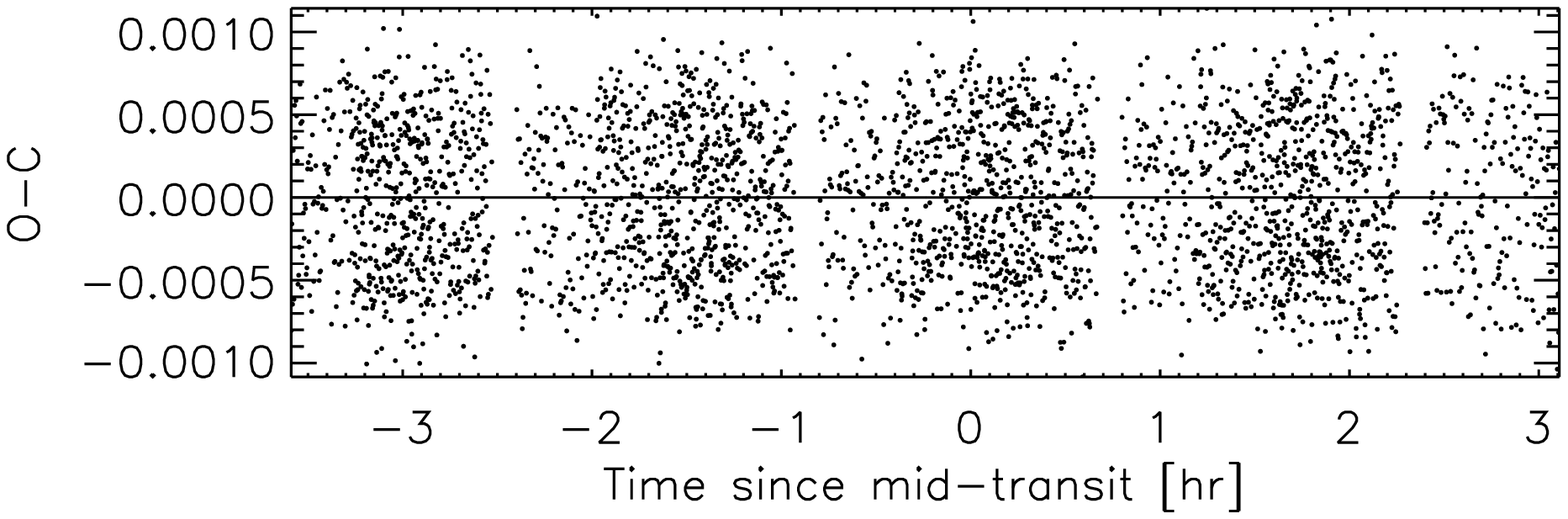}
 \caption{NICMOS transit light curve (1.1--2.0~$\mu$m) of HD~149026b.
   The data from 4 transits have been superimposed, after correcting
   for systematic effects. The solid line shows the best-fitting
   model.}
   \label{fig:comp}
\end{figure}

\begin{figure}[htbp] 
   \centering
   \epsscale{0.9}
 \plotone{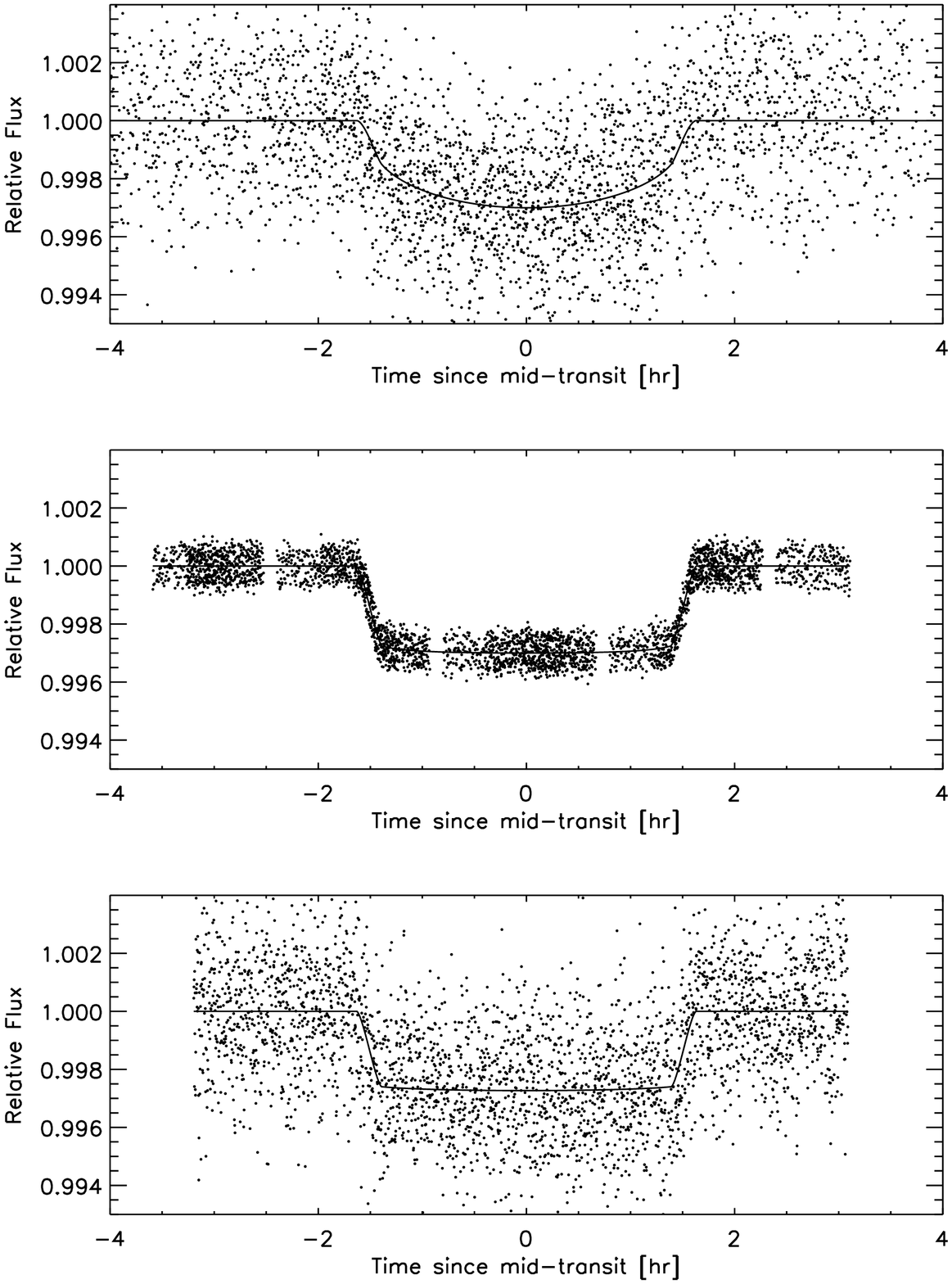}
 \caption{Comparison of the best available transit light curves of
   HD~149026.  Top panel: optical [Stromgren $(b+y)/2$] photometry
   from Sato et al.~(2005) and Winn et al.~(2008b), with a time
   sampling of 8.6~s and an rms residual of 2017~ppm. Middle
   panel: near-infrared [1.1--2.0~$\mu$m] photometry from this work,
   with a time sampling of 7.2~s and an rms residual of 440~ppm.
   Bottom panel: mid-infrared [8~$\mu$m] photometry from
   Nutzman et al.~(2008), with a time sampling of 7.4~s and an
   rms residual of 1854~ppm. The solid lines show the best-fitting
   model, which was calculated as described in \S~\ref{sec:joint}.}
   \label{fig:comp_binned}
\end{figure}

Plotted in Fig.~\ref{fig:rawperperts} is the measured flux after
dividing by the optimized function $f_{\rm sys}(t)$. This represents
our best effort to correct for the systematic effects. In
Fig.~\ref{fig:comp}, we show the results of combining the data from
all the visits (after correcting for systematic effects) into a single
transit light curve. In this composite light curve, the median time
between samples is $7.2$~s. Finally, in Fig.~\ref{fig:comp_binned} we
show a time-binned version of the composite light curve to allow a
visual comparison with the best previously-measured light curves, at
optical and mid-infrared wavelengths.


To determine the ``allowed range'' of each parameter---or, more
precisely, the {\it a posteriori}\, joint probability distribution of
all the parameter values---we employed the Markov chain Monte Carlo
(MCMC) technique (see, e.g., Winn et al.~2007; Burke et al.~2007). We
produced 8 chains of length $7.4 \times 10^6$ using a Gibbs sampler
and a Metropolis-Hastings jump condition such that each parameter had
an effective chain length of roughly $2\times 10^5$. This was
accomplished by adjusting the scale of the jump-function distribution
such that the probability of accepting a jump is roughly uniform and equal to approximately $40\%$ across all parameters. 
 To establish initial estimates
of parameter uncertainties, a preliminary Monte Carlo bootstrap
analysis was performed, based on the Levenberg-Marquardt least squares
minimization; then, the MCMC initial conditions were drawn from normal
distributions with widths equal to five times these initial error
estimates. The first $25\%$ of each chain was trimmed, and then all
the chains were concatenated. Each parameter had a Gelman \& Rubin
(1992) $R$ statistic smaller than 1.01, a sign of good convergence of
the posterior parameter distributions. For each parameter, the values
at each link of the chain were sorted. To describe the results, we
report the median (50\%) value, along with the interval between the
15.85\% and 84.15\% levels (the 68.3\% confidence interval). The
results are given in Column 2 of Table~(\ref{tab:MCMCout}).

\subsection{Results from NICMOS photometric analysis} \label{sec:photo_results}

For the orbital inclination, we find $i = 84.55^{+0.35}_{-0.81}$~deg.
This is in agreement with the independent estimate of $i =
85.4^{+0.9}_{-0.8}$~deg by Nutzman et al.~(2008), using the 8~$\mu$m
channel of the Infrared Array Camera (IRAC) aboard the {\it Spitzer
  Space
  Telescope}. For the normalized orbital distance, we find $a/R_\star
= 6.01^{+0.17}_{-0.23}$. This is also in agreement with the results of
Nutzman et al.~(2008), who found $a/R_\star = 6.20^{+0.28}_{-0.63}$.
The new result is more precise, which (as we will show in
\S~\ref{sec:stellar}) leads to tighter constraints on the stellar mass
and radius. This is important because the uncertainties in the stellar
properties have been the limiting factors in the analysis of this
system. Based on the preceding results, we find the impact parameter
(defined as $b = a\cos i/R_\star$) to be $0.571^{+0.044}_{-0.038}$.
This is the tightest such constraint that has been achieved for HD~149026b. The
earliest measurements of the impact parameter were consistent with
zero (Sato et al.~2005, Charbonneau et al.~2006, Winn et al.~2008b), a
situation leading to strong degeneracies among the transit parameters
(Carter et al.~2008). More recently, Nutzman et al.~(2008) found $b =
0.62^{+0.08}_{-0.24}$, consistent with the new and more precise
result. The increased precision in $a/R_\star$ and $b$ is a direct
consequence of the improved precision with which the ingress (and
egress) duration is known (Carter et al.~2008). In this sense, the
greatest value of the NICMOS data is in the good coverage of the
ingress and egress phases.

The enhanced precision of the NICMOS data relative to previous data
sets does not lead to a correspondingly enhanced precision in the
planet-to-star radius ratio.  This is because we allowed the time
series from each orbit to have its own adjustable flux
multiplier. Consequently, all of the information about $R_p/R_\star$
comes only from those orbits that span an entire ingress or egress
event. However, the value of $R_p/R_\star$ that we derive is at least
comparable in precision to previous determinations. We find
$R_p/R_\star = 0.05416^{+0.00091}_{-0.00070}$. Interestingly this is
larger by 2$\sigma$ than the values derived previously, which were
based on optical and mid-infrared data. Winn et al.~(2008b) found
$R_p/R_\star = 0.0491^{+0.0018}_{-0.0005}$ based on Stromgren
$(b+y)/2$ photometry, and Nutzman et al.~(2008) found $R_p/R_\star =
0.05158\pm0.00077$ based on 8~$\mu$m photometry.

\begin{figure}[htbp] 
   \centering
   \epsscale{0.8}
   \plotone{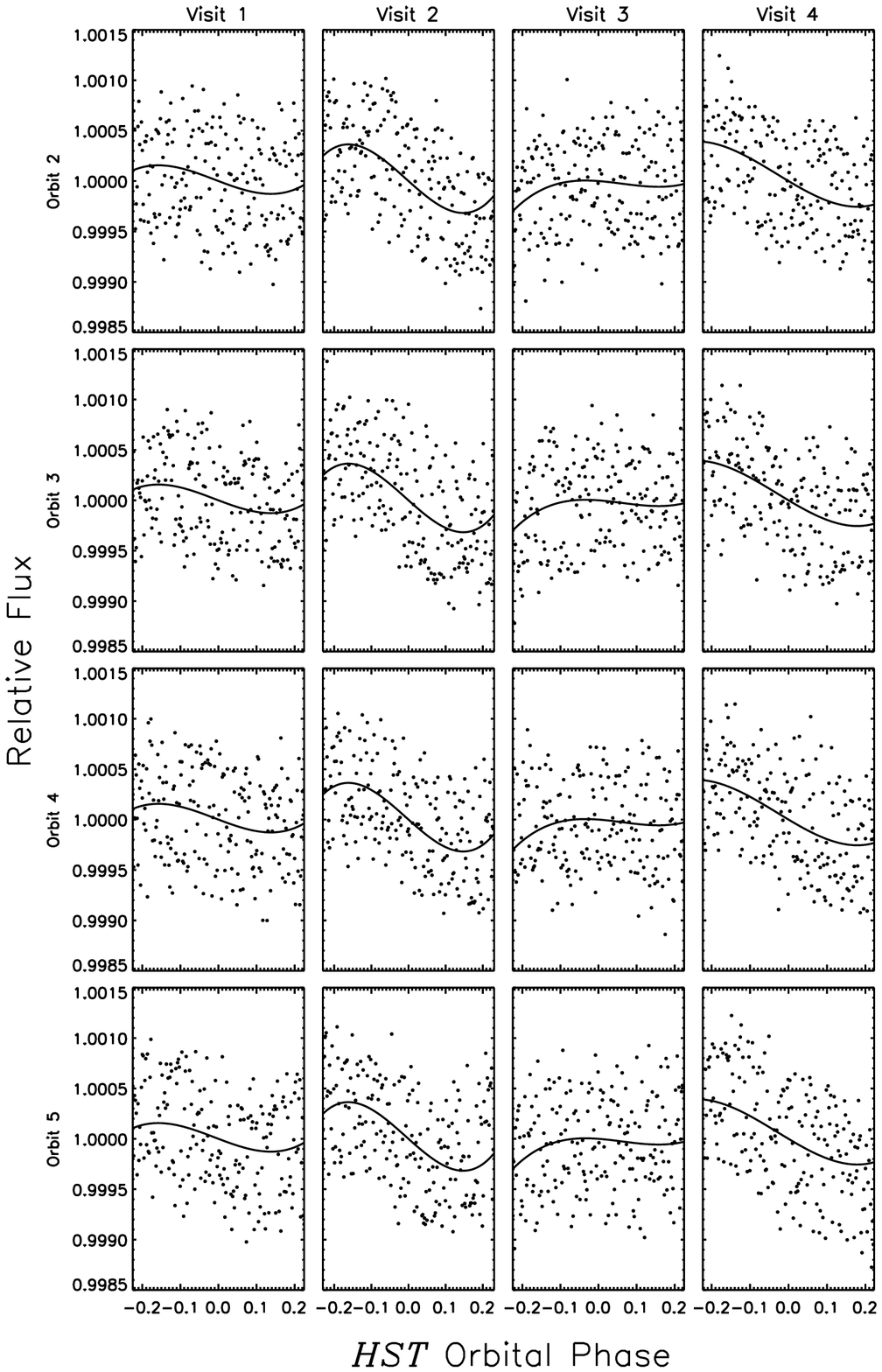}
   \caption{Isolation of the intra-orbital variations. The flux time
     series has been divided by the optimized flux multipliers $f_o^v$
     and by the optimized transit model. The remaining variation
     appears to present a consistent pattern among all orbits within a
     given visit, as assumed in our model. The solid line is the
     optimized model. Each column shows data from all orbits of a given visit.  
      Each row shows orbits arranged from first to last in rows from top to bottom, respectively.}
   \label{fig:phase}
\end{figure}

Since we do not understand all of our noise sources with a
physically-grounded model, we cannot rule out the possibility that the
discrepancy between our result and the previous results is due to a
faulty model of the systematic effects. The culprit would probably
need to be the polynomial function of orbital phase. We do find that
our result is unaffected if we replace the polynomial function of
$\phi$ with trigonometric functions, as mentioned previously; and we
also find that our results are unchanged if we use a linear limb darkening law
 or fix the quadratic limb darkening parameters to those tabulated by Claret (2000). 
 These tests do not prove that our results are valid
but they do suggest that our analysis procedure is robust to changes
in the functional form of the model. However, to the extent that the
intra-orbital variations are not strictly repeatable within a given
visit, our model would produce biased results. Fig.~\ref{fig:phase}
shows the data after dividing by the optimized values of $f_o^v$ (the
orbit-specific flux multipliers) and dividing by the optimized transit
model. The purpose of the divisions is to isolate the intra-orbital
systematic effects, which do appear consistent among the orbits within
a given visit.

Another possibility for the discrepancy in the transit depth between
our near-infrared result and the previous optical and mid-infrared
results is differential absorption due to constituents in the outer,
optically-thin portion of the planet's atmosphere. This interpretation
is the basis of the ``transmission spectroscopy'' technique for
identifying constituents of exoplanetary atmospheres pioneered by
Charbonneau et al.~(2002). We consider this possibility at some length
in sections \ref{sec:joint} and \ref{sec:conclusions}.

\begin{figure}[htbp] 
   \centering
   \epsscale{1.00}
   \plotone{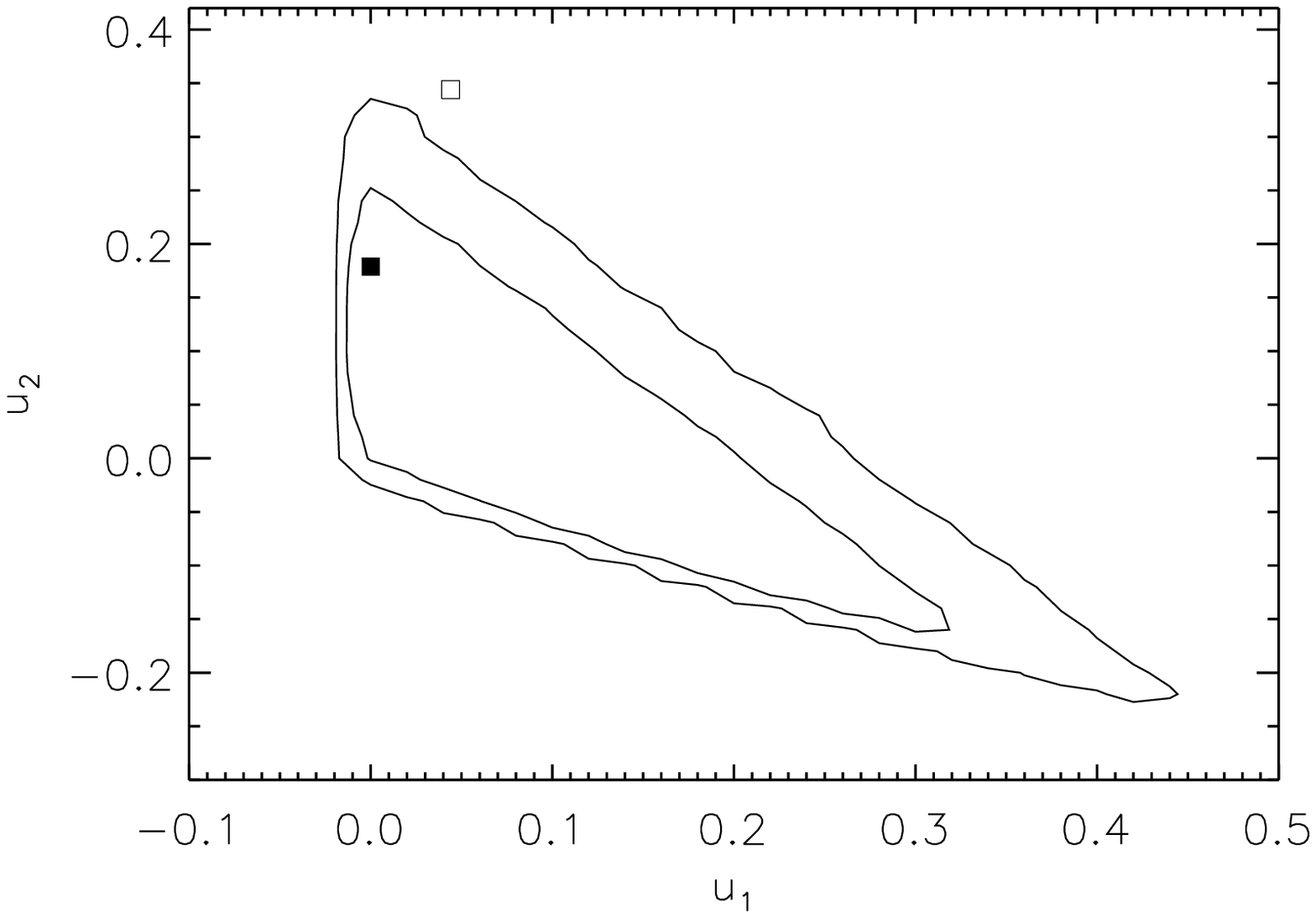}
   \caption{Results for the limb-darkening parameters $\UA$ and $\UB$.
     The contours are the $68\%$ and $95\%$ confidence regions as
     determined by the MCMC analysis of the photometric time series.
     The solid square is the minimum-$\chi^2$ solution, $\UA = 0.0$,
     $\UB = 0.1789$.  The open square marks the tabulated values of
     Claret~(2000) for $H$ band ($\UA = 0.044$, $\UB = 0.344$). }
   \label{fig:limb}
\end{figure}

We find that the center-to-limb variation is less pronounced than was
expected based on the tabulated limb darkening coefficients of
Claret~(2000). Fig.~\ref{fig:limb} shows the confidence contours in
the $\UA$, $\UB$ parameter space. The open square corresponds to the
tabulated values for $H$ band (for a star with $T_{\rm eff} = 6250~K$,
$\log g_\star = 4.5$, $[{\rm Fe}/{\rm H}] = 0.3$, matching the
properties of HD~149026). The tabulated values are excluded with
$>$95\% confidence.

Two quantities intrinsic to the star and planet that may be calculated
directly in terms of observables are the surface gravity of the
planet, and the mean density of the star.  The surface gravity of the
planet is calculated as (Southworth et al.~2007)
\begin{eqnarray}
	g_p & = & \frac{2 \pi}{P} \frac{K}{(R_p/a)^2 \sin i},
\end{eqnarray}
where $K$ is the semiamplitude of the stellar radial-velocity
variation ($43.3\pm 1.2$~m~s$^{-1}$; Sato et al.~2005), and $R_p/a$
and $\sin i$ are derived from our MCMC analysis. We find $\log g_p =
3.132_{-0.035}^{+0.029}$ where $g_p$ is in cgs units. The mean stellar
density $\rho_\star$ is calculated as (Seager \& Mallen-Ornelas 2003,
Sozzetti et al.~2007)
\begin{eqnarray}
	\rho_{\star} =
   \frac{3 \pi}{GP^2}
   \left(\frac{a}{R_\star}\right)^3 - 
   \rho_p \left(\frac{R_p}{R_\star}\right)^3 \label{eq:dens}
\end{eqnarray}
where $\rho_p$ is the mean density of the exoplanet.  We may neglect
the correction term involving the planetary density as $\rho_\star
\sim \rho_p$, $(R_p/R_\star)^3\sim1\times10^{-4}$ and the fractional
uncertainty in $a/R_\star$ is larger than the fractional uncertainty
in $R_p/R_\star$ by a factor of 2.

\section{Stellar Parameters} \label{sec:stellar}

The basic inputs to models of the planetary interior are the planetary
mass $M_p$ and radius $R_p$, in units of grams and kilometers, or in
units of Jupiter's mass and radius. Transit photometry and Doppler
velocimetry alone do not determine these quantities. Additional
information about the star must be introduced. Several techniques for
estimating the stellar mass $M_\star$ and radius $R_\star$ were
reviewed by Winn et al.\ (2008b). We chose to estimate $M_\star$ and
$R_\star$ using stellar-evolution models that are constrained by the
best available, relevant, observable properties of the star: the mean
density $0.497_{-0.057}^{+0.042}$~g~cm$^{-3}$ determined from our
light-curve analysis, the absolute magnitude $M_V = 3.65 \pm 0.12$
derived from the {\em Hipparcos}\, parallax and apparent magnitude
[$\pi = 12.59 \pm 0.70$ mas, $V = 8.15\pm0.02$; van Leeuwen~(2007)], 
effective temperature [$T_{\rm eff} = 6160\pm 50$~K, a
weighted mean of the results from Sato et al.~(2005) and Masana et
al.~(2006)], and metallicity [$0.36\pm0.08$, from Sato et al.~(2005)
with a more conservative error bar]. We chose not to use the
spectroscopically-determined stellar surface gravity [$\log g_\star=
4.26\pm0.07$; Sato et al.~(2005)] because the
photometrically-determined value of $\rho_\star$ provides an
effectively tighter constraint, and because
spectroscopically-determined surface gravities have been found to be
susceptible to systematic error (see, e.g., Winn et al.~2008a, \citet{2008ApJ...677..657J}).

\begin{figure}[htbp] 
   \centering
   \epsscale{1.00}
   \plotone{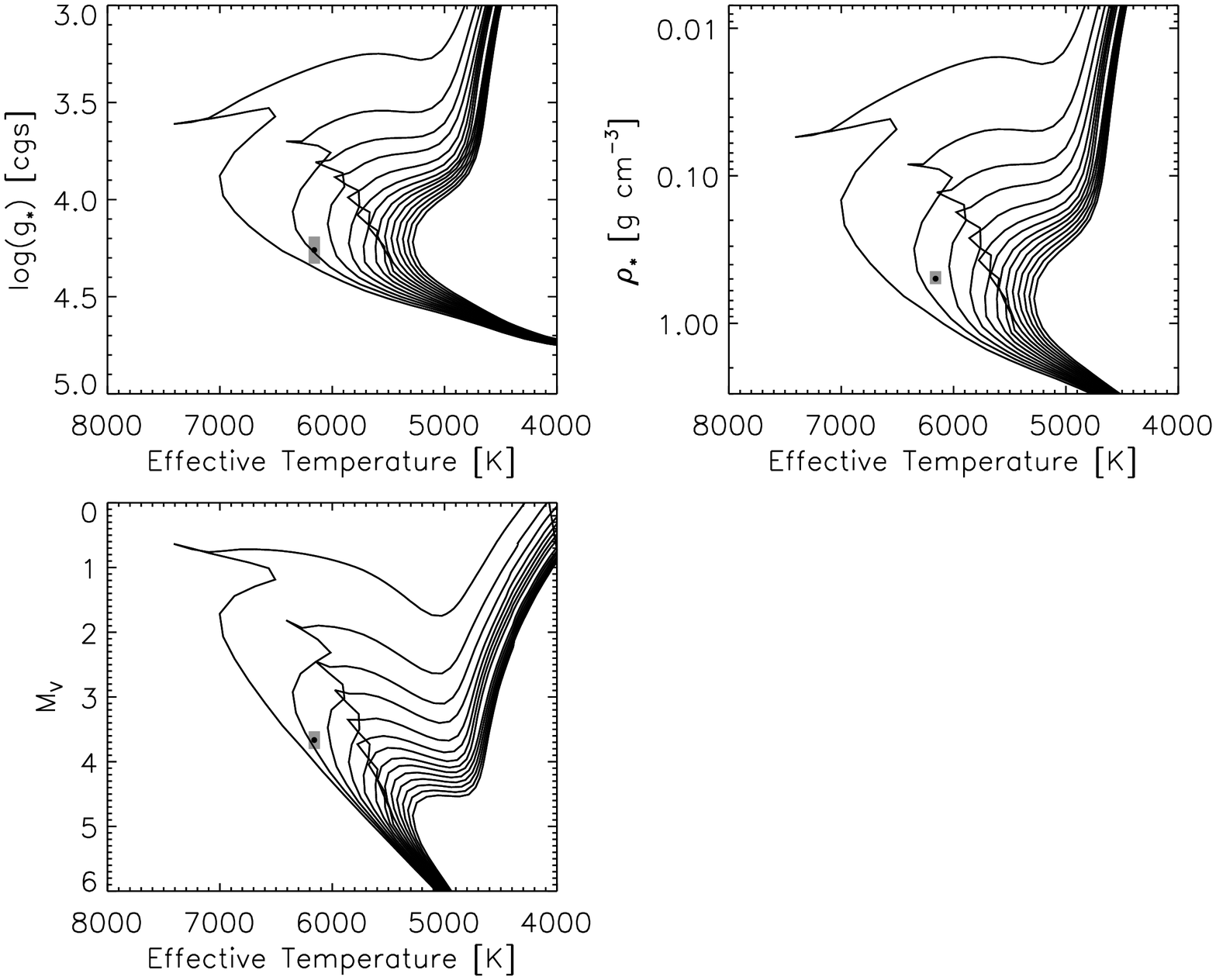}
   \caption{Stellar-evolutionary model isochrones, from the
     Yonsei-Yale series by Yi et al.~(2001). The points and shaded
     boxes represent the observationally-determined values and
     1$\sigma$ errors. Here, surface gravity is determined
     spectroscopically (Sato et al.~2005), $M_V$ is derived from {\em
       Hipparcos} parallax and $V$ magnitudes, and $\rho_\star$ is
     determined photometrically from the transit light curve.
     Isochrones are shown for ages of 1 to 13 Gyr (from left to right)
     in steps of 1 Gyr for a fixed stellar metallicity of
     [Fe/H]$=0.36$. }
   \label{fig:isochrones}
\end{figure}

Following the procedure of Torres, Winn, \& Holman (2008), we employed
Yonsei-Yale stellar models\footnote{
	We chose the Y$^2$ models mainly because of the convenient form in
which they are publicly available. Other stellar-evolutionary models
are available, and other investigators have examined the sensitivity
of results such as ours to the choice of model. For HD~149026 in
particular, Southworth (2008) found that the Y$^2$ models and
independent models by Claret~(2007) gave results for $M_\star$ and
$R_\star$ that agreed to within 1\%. Another set of publicly available
models, the Padova models of Girardi et al.~(2000), are not computed
for the high metallicity observed for HD~149026; but at zero
metallicity, at least, both Torres et al.~(2008) and Southworth~(2008)
found that the Padova models give results for the stellar mass and
radius that are also within 1\% of the Y$^2$ results.
}	(Yi et al.~2001, Demarque et
al.~2004). Model isochrones were interpolated in both age and
metallicity, for metallicities [Fe/H] ranging from 0.28 to 0.43 and
for ages ranging from $0.1$ to $14$~Gyr, in steps of
$0.1$~Gyr. Fig.~\ref{fig:isochrones} shows several of these
theoretical isochrones, along with some of the observational
constraints. The upper left panel illustrates the constraint due to
the spectroscopically-determined surface gravity, even though we did
not actually apply that constraint, as explained above. It is evident
that the constraint due to $\rho_\star$ is stronger.

The isochrones were interpolated to provide a fine grid in stellar
mass (with a step size of $0.005~M_{\odot}$). We then assumed that the
likelihood of each point on the interpolated isochrones was
proportional to $\exp(-\chi_\star^2/2)$, where
\begin{eqnarray}
  \chi_*^2 =
 \left(\frac{\Delta {\rm [Fe/H]}}{\sigma_{{\rm[Fe/H]}}}\right)^2 + 
 \left( \frac{\Delta T_{\rm eff}}{\sigma_{T_{\rm eff}}}\right)^2 + 
 \left(\frac{\Delta M_V}{\sigma_{M_V}}\right)^2 + 
 \left( \frac{\Delta \rho_\star}{\sigma_{\rho_\star}}\right)^2,
\end{eqnarray}
and the $\Delta$ quantities denote the differences between the
observed and calculated values.  The asymmetry in the error
distribution for $\rho_\star$ was taken into account. Additionally,
the likelihood was taken to be proportional to an Salpeter initial
mass function, $\xi(M) \propto M^{-(1+x)}$ with $x=1.35$
(Salpeter~1955). The joint probability function, $\JPF$, was
taken to be proportional to the likelihood, viz.,
\begin{eqnarray}
 \JPF(R_\star, M_\star, T_{\rm eff}, \log g_\star, M_V, {\rm [Fe/H]}, \rho_\star, {\rm Age}) \propto \exp(-\chi_*^2/2).
\end{eqnarray}
For a given parameter $X_0$ from this list, we calculated the
cumulative distribution function (CDF) by numerically evaluating
\begin{eqnarray}
  \CDF(x) = \int_{-\infty}^x~dX_0 \int_{-\infty}^\infty~dX_1 \cdots \int_{-\infty}^\infty~dX_N~~
  \JPF(X_0, X_1, \cdots, X_N).
\end{eqnarray}
For each parameter, we record the values of $x$ for which the CDF
takes the values 15.85\%, 50\%, and 84.15\%. The 50\% level (the
median) is reported as the ``best-fit value'' and the interval between
the 84.15\% and 15.85\% levels is reported as the $68.3\%$ (1$\sigma$)
confidence interval.\footnote{Although our procedure was inspired by the
  work of Torres et al.~(2008) and is similar in almost all respects,
  there is one significant difference. The best fit values reported by Torres et al.~(2008) 
  were those that minimized $\chi^2$ as in our analysis.  The difference is that Torres et al.~(2008) estimated
  the 1$\sigma$ errors in the stellar properties based on the total
  span of the calculated values that gave agreement within 1$\sigma$
  with the observables.  Effectively, they assumed a uniform error
  distribution for each observable, rather than a Gaussian error
  distribution as we have done.  Consequently, our method produces
  smaller error intervals in the stellar properties. Caution would
  dictate that larger error intervals are desirable, especially since
  we are relying on the theoretical isochrones that surely have some
  unaccounted-for systematic errors. However, using a uniform error
  distribution for the observables is an arbitrary way to inflate the
  output errors, and the true error distribution for the observables
  is probably closer to Gaussian. For these reasons we chose our
  approach and emphasize the caveat that our results place complete
  trust in the Y$^2$ isochrones.}

Table~(\ref{tab:MCMCout}) reports the best-fit stellar parameters and
confidence intervals.  We find the stellar radius to be $R_\star =
1.541^{+0.046}_{-0.042} ~R_\sun$. This is larger than (but in
agreement with) the previous estimates of $R_\star = 1.46\pm0.10
~R_\sun$ by Sato et al.~(2005), and $R_\star = 1.497\pm0.069 ~R_\sun$)
by Nutzman et al.~(2008). By combining the derived distribution for
$R_\star$ with the photometrically-determined distribution for
$R_p/R_\star$, we find the planetary radius to be $R_p =
0.813^{+0.027}_{-0.025}~R_{\rm Jup}$. This is larger than any previous
result. Using optical photometry, Sato et al.~(2005) found
$0.725\pm0.050 ~R_{\rm Jup}$, Charbonneau et al.~(2006) found
$0.726\pm0.064 ~R_{\rm Jup}$, and Winn et al.~(2008b) found
$0.71\pm0.05~R_{\rm Jup}$. Using mid-infrared photometry, Nutzman et
al.~(2008) found $0.755\pm0.040 ~R_{\rm Jup}$. It is important to note
that these determinations were not wholly independent, and therefore
should not be combined into a weighted average. They all used many
common inputs for the stellar properties, and the analyses of optical
photometry all included a common subset of at least 3 light curves.

As mentioned previously, we did not apply any constraint to the models
based on the spectroscopically determined value of $\log g_\star$.
However, given our results for $M_\star$ and $R_\star$ we computed the
implied value of $\log g_\star$, finding $\log g_\star =
4.189^{+0.020}_{-0.021}$. This in agreement with, and is more precise
than, the spectroscopically-determined value of $\log g_\star =
4.26\pm0.07$ (Sato et al.~2005).

\section{Joint Analysis with Optical and Mid-Infrared Light Curves} \label{sec:joint}

Transit observations of HD~149026b have now been made at optical
wavelengths (Sato et al.~2005, Charbonneau et al.~2006, Winn et
al.~2008b), near-infrared wavelengths (this work), and mid-infrared
wavelengths (Nutzman et al.~2008). In this section we repeat our
analysis on all of these data, in order to bring all of these data to
bear on the determination of the system parameters, while seeking
possible wavelength variations in the planet-to-star radius ratio.

It is reasonable to require consistency across these data in the
parameters relating to the orbital configuration of the transit, such
as the inclination angle and normalized semi-major axis. However, the
inferred planet-to-star radius ratio is a wavelength-dependent
quantity, depending on the opacity of the exoplanetary atmosphere and
the emergent flux from the planetary nightside (which is expected to
be unimportant).  With this in mind, we performed a joint analysis of
all of the data, requiring consistency in $\cos i$ and $a/R_\star$ but
allowing $R_p/R_\star$ to take separate values for each of the three
types of data: optical, near-infrared, and mid-infrared.

Specifically, we fitted our NICMOS data, the 8~$\mu$m IRAC time series
of Nutzman et al.~(2008), and the 8 light curves obtained in the
Str\"{o}mgren $(b+y)/2$ band by Sato et al.~(2005) and Winn et
al.~(2008b). Our photometric model for the NICMOS data, including the
associated systematic effects, has already been described. For the
data sets presented by other authors, we followed those authors'
prescriptions to account for systematic errors. For the $(b+y)/2$
data, we corrected the data by allowing the out-of-transit flux to be
a linear function of time. For the 8~$\mu$m data, we modeled the
time-variable sensitivity of the detector (the ``ramp'') as a
multiplicative correction, $f_{\rm
  sys} = a_0 + a_1 \log(t-t_0) + a_2 \log^2(t-t_0)$, where $t_0$ is
the time immediately prior to the start of the observation.

We performed an MCMC analysis of this joint data set. The free
parameters relating to the transit model were the three values of
$R_p/R_\star$ (corresponding to the ratios measured at approximately
0.5~$\mu$m, 1.5~$\mu$m, and 8.0~$\mu$m); the geometric parameters
$\cos i$ and $a/R_\star$; the quadratic limb-darkening coefficients
for the NICMOS light curve; the linear limb-darkening coefficients for
the optical and infrared light curves (for which the precision of the
data do not justify the more accurate quadratic law); and the
mid-transit times for the NICMOS data and the IRAC data.\footnote{To
  keep the number of
  parameters as small as possible, we did not vary the optical
  mid-transit
  times or baseline correction parameters at this stage, having found
  that they are uncorrelated with the other parameters of interest.}
We also fitted for the ramp-correction terms for the IRAC data and
the parameters relating to the flux offsets and intra-orbital
variations for the NICMOS data. Six chains of length $9\times10^6$
were created, representing approximately $2\times 10^5$
correlation-lengths per parameter. These were concatenated after
removing the first $25\%$ of each chain. The Gelman-Rubin $R$
statistic was smaller than 1.01 for each parameter. We then repeated
the analyses that were described in \S~\ref{sec:nicanalysis} and
\S~\ref{sec:stellar} to determine the stellar, planetary, and orbital
parameters, based on this joint analysis. The results are tabulated in
Table~(\ref{tab:MCMCout}).

The results for the geometric parameters are hardly changed from the
NICMOS-only analysis, a reflection of the greater precision of the
NICMOS light curve. The planet-to-star radius ratio was found to be
larger for the NICMOS data than for the other bandpasses, as was
already evident from the comparison of our NICMOS-only analysis to
previously published analyses. Fig.~\ref{fig:transspec} shows the
variation in $(R_p/R_\star)^2$ with wavelength. The quantity
$(R_p/R_\star)^2$ is essentially the transit depth, or fractional loss
of light during the total phase of the transit, after ``removing'' the
effects of limb darkening. The radius ratios that we derive for the
mid-infrared and optical data are in agreement with those reported
previously. For the IRAC data we find $R_p/R_\star =
0.05188_{-0.00086}^{+0.00084}$ as compared to the value
$0.05158\pm0.00077$ reported by Nutzman et al.~(2008). For the
$(b+y)/2$ data we find $R_p/R_\star = 0.05070_{-0.00088}^{+0.00058}$
as compared to the value $0.0491^{+0.0018}_{-0.0005}$ found by Winn et
al.~(2008b). The precision in the optical $R_p/R_\star$ has been
increased because the NICMOS data pins down all of the other
parameters that are correlated with $R_p/R_\star$.

\section{Ephemeris and transit timing} \label{sec:timing}

\begin{figure}[htbp] 
   \centering
      \plotone{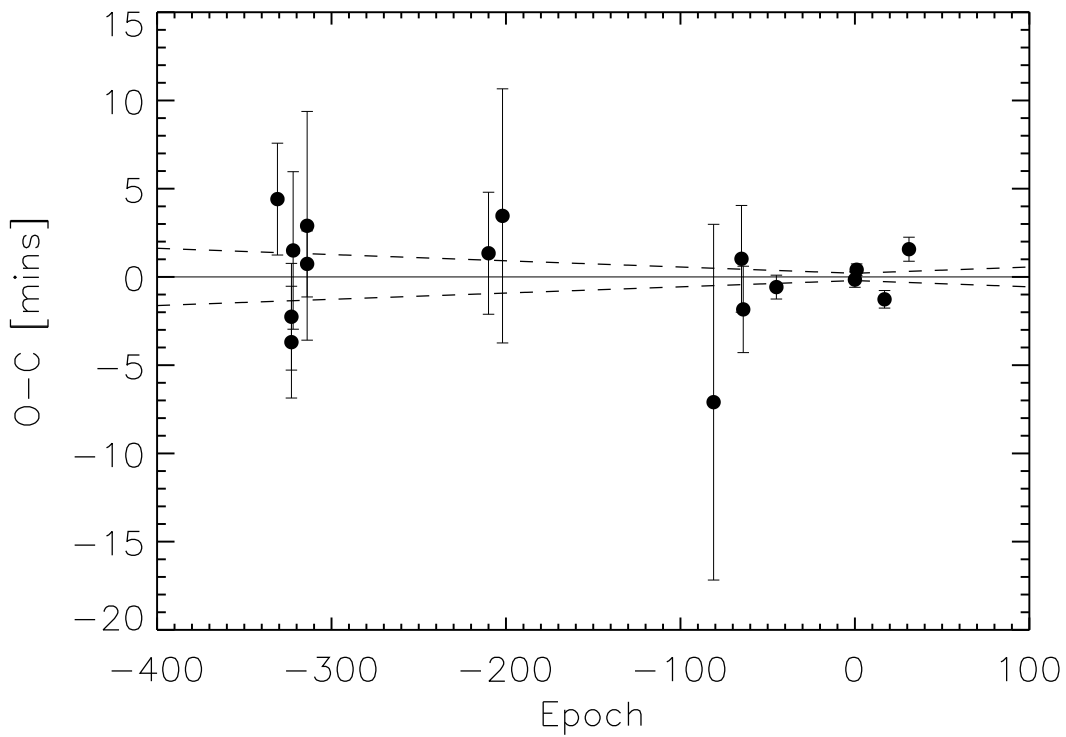}
      \caption{Transit-timing variations for HD~149026b. The differences
        between observed and calculated mid-transit times are plotted.
        The last 4 points represent the new NICMOS observations.
        The dashed
        lines show the 1$\sigma$ range in the calculated times
        according to the linear ephemeris presented in
        \S~\ref{sec:timing}.}
   \label{fig:timing}
\end{figure}

The NICMOS-only analysis resulted in the measurement of four distinct
mid-transit times, with uncertainties smaller than 45~s. These are
given in Table~(\ref{tab:times}). We pooled together all of the
independent measurements of mid-transit times from Winn et al.~(2008b),
Nutzman et al.~(2008), and this work, to derive a new transit
ephemeris. We fitted the times to a linear function of the integral
epoch $E$,
\begin{eqnarray}
	T_c(E) = T_c(0)+E P
\end{eqnarray}
where $P$ is the period and $T_c(0)$ is the mid-transit time at some
fiducial epoch. The results were $T_c(0) = 2454456.78761\pm0.00014$
HJD and $P = 2.8758911\pm0.0000025$ days. The linear fit had $\chi^2 =
20.16$ and 14 degrees of freedom. This is a marginally unacceptable
fit. The formal probability to find a value of $\chi^2$ this large is
15\%. Further transit observations are needed to distinguish the
possibilities of a genuine period variation, a statistical fluke, and
underestimated timing errors. Fig.~\ref{fig:timing} shows the O--C
(observed minus calculated) timing diagram.

\section{Discussion} \label{sec:conclusions}

We have presented observations of four transits of HD~149026b at
near-infrared wavelengths with the {\it HST}\, NICMOS detector. The
NICMOS data place the strongest constraints yet on the geometrical
system parameters. In particular, the increased precision of the
measurement of the normalized semi-major axis ($a/R_\star$) leads to
an improved estimate of the mean stellar density, which was then
coupled with stellar evolution models to constrain the stellar mass
and radius. Improved knowledge of the stellar mass and radius leads to
greater precision in the planetary mass and radius. We have found a
larger stellar radius, and a larger planet-to-star radius ratio, than
previous estimates. As a result of these two factors, we have also
found the planetary radius to be larger than previously thought. The
planet has ``grown'' by about 7\%.

Despite this increase, our results are still consistent with the
contention that HD~149026b is highly enriched in heavy elements. It is
still smaller than expected for a hydrogen-helium planet with the
given mass and degree of stellar irradiation (Burrows et al.~2007).
For comparison, the tabulated models by Fortney et al.~(2007) predict a $1.3~R_{\rm Jup}$ 
hydrogen-helium HD~149026b at an age of $1$~Gyr.  A variety
of models have been developed to estimate the heavy-element content of
HD~149026b (Sato et al.~2005, Fortney et al.~2006, Ikoma et al.~2006,
Burrows et al.~2007), most of which suppose that the metals are
confined to an inner core of material beneath a hydrogen-helium
envelope. Other physical considerations in these models include
the equation of state for heavy elements at core pressures,
atmospheric opacities and the upper boundary condition where energy is
delivered from the star. To determine a revised estimate for the
heavy-element content, we used the tabulated models provided by Sato
et al.~(2005) and Fortney et al.~(2007), and interpolated the
tabulated results as appropriate for the planetary radius, planetary
mass, and degree of irradiation that follow from the parameters
determined from the NICMOS data. We find a core mass in the range of
$45-70$~$M_\earth$, depending on assumed stellar age and core density. Thus,
the interpretation of the planet as highly enriched is unaffected,
although the required amount of enrichment is slightly reduced.

It is also interesting that the planet-to-star area ratio,
$(R_p/R_\star)^2$, was found to be 2--3$\sigma$ larger in the NICMOS
band (1.1--2.0~$\mu$m) than in the optical band (0.45--0.55~$\mu$m) or
mid-infrared bands (6.5--9.5~$\mu$m), while the results for the latter
two bands are in agreement. Caution dictates that this discrepancy
should not be over-interpreted. It is possible that the discrepancy is
at least partly the result of unresolved systematic errors in any of
the data sets. We have already noted that the noise in the NICMOS data
exceeds the photon noise level by a factor of 2, and is not well
understood.

However, it is also worth considering that this wavelength-dependent
variation represents selective absorption by constituents in the
outermost layer of the planet's atmosphere. Molecules with strong
absorption bands at near-infrared wavelengths would cause the transit
to appear deeper at those wavelengths. Strong bands are expected for
the common molecules CO, H$_2$O, and CH$_4$ (Brown~2001, Hubbard et
al.~2001, Seager \& Sasselov~2000). If this were the case, then a
detailed analysis of the NICMOS spectrophotometry---breaking it down
into smaller wavelength bins, as opposed to summing the entire
first-order spectrum---might be used to identify some constituents of
the planet's atmosphere. In addition, more care would be needed in
choosing which radius to use in the comparison with models of the
planet's interior. It is beyond the scope of this paper to analyze the
wavelength dependence of the transit depth across the NICMOS band, or
to compute a realistic atmospheric model to see if the contrast
between the optical, near-infrared, and mid-infrared results can be
accommodated. We can, however, perform an order-of-magnitude
calculation to check on the plausibility of this interpretation.

\begin{figure}[htbp]
   \centering
   \plotone{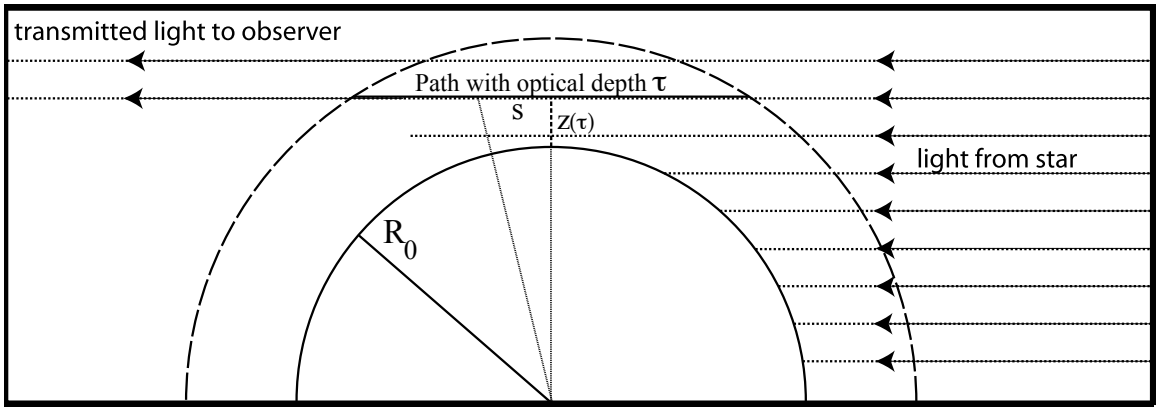}
   \caption{Illustration of wavelength-dependent absorption.
     Shown are some rays that skirt the planetary atmosphere
     on their way to Earth.
     At some height $z(\tau)$ above the fiducial radius $R_0$,
     the ray (parameterized by path length $s$) has an optical depth of $\tau$ (solid line).  
     Light that follows paths with $z < z(\tau = 1)$ are
     mainly absorbed. The height $z(\tau = 1)$ corresponding to optical depth of unity
     depends on wavelength, giving rise to a wavelength-dependent
     transit radius.}
   \label{fig:atms}
\end{figure}

\begin{figure}[htbp] 
   \centering
   \plotone{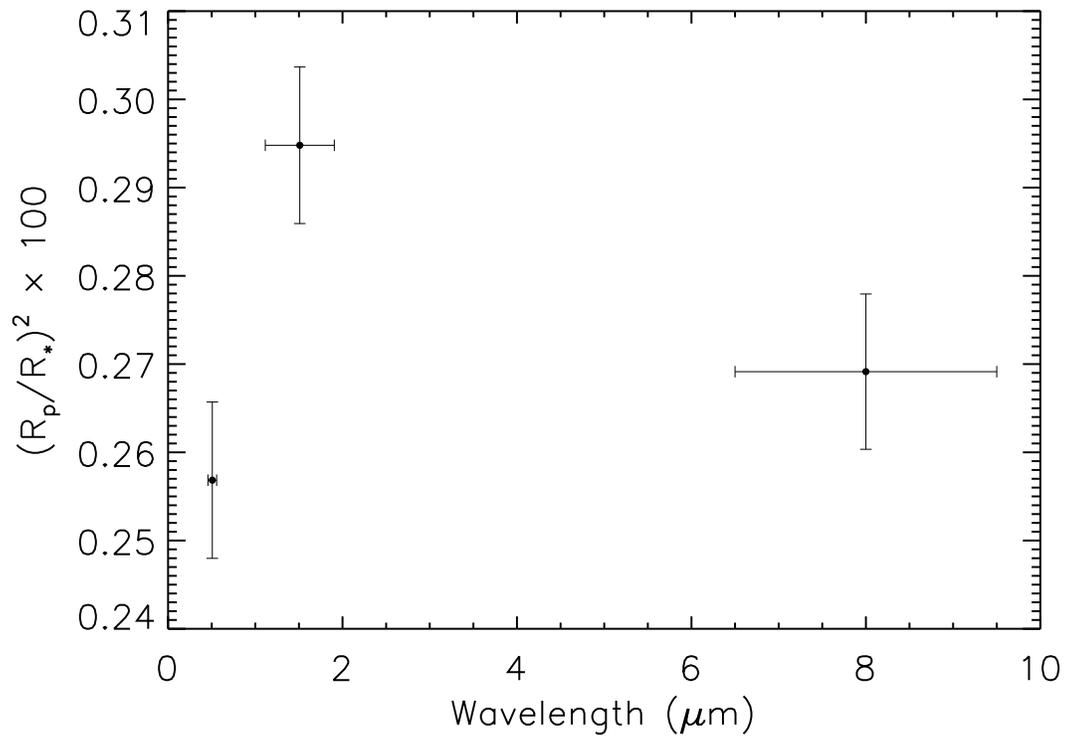}
   \caption{The planet-to-star area ratio, $(R_p/R_\star)^2$, as
     a function of observing wavelength, based on a joint fit
     to the NICMOS data, the $(b+y)/2$ data of Winn et al.~(2008),
     and the IRAC data of Nutzman et al.~(2008).
     The horizontal error bars show the
     approximate wavelength range of each bandpass.}
   \label{fig:transspec}
\end{figure}

Let $z(\tau)$ be the height in the planet's atmosphere at which the optical
depth is $\tau$ for a path from the star to the observer, as diagrammed
in Fig.~\ref{fig:atms}. This height is measured relative to an
atmospheric base radius $R_0$, where the planet is optically thick at
all relevant wavelengths. The height $z(\tau)$ depends, in part, on the
wavelength-dependent opacity and the density profile of the
atmosphere. We define $R_p(\lambda)$ as $R_0 + z(\tau = 1)$ and $\DEP
\equiv (R_p/R_\star)^2$. If we assume that $z(1) \ll \RA$, then the
transit depth is approximately linear in $z(1)$:
\begin{eqnarray}
  \DEP & = & \frac{[\RA+z(1)]^2}{R_\star^2} \approx
     \left(\frac{\RA}{R_\star}\right)^2
     \left[1+\frac{2z(1)}{\RA}\right].
\end{eqnarray}

Next, we consider the difference in $\DEP$ as measured in two distinct
wavelength bands:
\begin{eqnarray}
  \DEP_1-\DEP_2 &\approx& \frac{2\RA}{R_\star^2} \left[z_1(1)-z_2(1)\right]
   = \frac{2\RA}{R_\star^2} \delta z \label{eqn:DF}
\end{eqnarray}
where we have defined the height difference $\delta z \equiv z_1(1)-z_2(1)$.
The height difference $\delta z$ reflects differing levels of
absorption in the two bands. Solving for $\delta z$ in
Eqn.~(\ref{eqn:DF}) we find
\begin{eqnarray}
\delta z & = & \frac{1}{2} \left[ \frac{\DEP_1-\DEP_2}{(\RA/R_\star)^2}\right] \RA.
\label{eq:dz}
\end{eqnarray}

For HD~149026b, to evaluate the idea that the larger near-infrared
measurement of $\DEP$ is due to molecular absorption, we assume that
$R_0$ is the optically-derived radius ($\RA = 0.757~R_{\rm Jup}$)
because the optical spectrum is expected to show comparatively weak
absorption features (Brown 2001, Seager \& Sasselov 2000). Using
Eqn.~(\ref{eq:dz}), the $0.035\%$ difference in $\DEP$ between
near-infrared and mid-infrared wavelengths implies $\delta z \approx
2500$~km ($4\%$ of $R_0$). To judge if this is realistic, we are
interested in expressing $\delta z$ in units of the pressure scale
height, for which an order-of-magnitude expression is $\PH = k T/\mu
m_p g_p$, where $T$ is a representative atmospheric temperature, $\mu$
is the mean molecular weight of the atmosphere, $m_p$ is the proton
mass and $g_p$ is the surface gravity. Using the surface gravity that
was determined from our analysis of the optical light curve ($g_p =
1535$ cm~s$^{-2}$), and assuming a H$_2$--He atmosphere with $T =
2300$~K [as measured at 8~$\mu$m by Harrington et al.~(2007)], we find
$\PH \approx 530$ km, and $\NS \approx 5$. If instead we use the
planet's predicted temperature at thermal equilibrium with the
incident stellar radiation ($T=1700$~K), we find $\PH\approx 400$ km
and $\NS \approx 6$.

If we assume further that the absorbers have an exponential density
profile,
\begin{eqnarray}
  \rho(z) &  = & \rho(0) \exp\left(-\frac{z}{\Dh}\right),
\end{eqnarray}
where $h$ is the density scale height, then we may express $z(\tau)$ in
terms of the opacity $\sigma$ of the absorbing molecules, as follows.
By integrating the optical depth $\tau$ across the optical path at
height $z \ll R_0$ (as illustrated in Fig.~\ref{fig:atms}), where a change of path length $ds$ results in a change in optical depth $d\tau$ as
\begin{eqnarray}
	d\tau = -s \sigma \rho(0) \frac{\exp\left[-\frac{R_0}{h}\left(\sqrt{\left(1+z/R_0\right)^2+(s/R_0)^2}-1\right)\right] }{\sqrt{\left(1+z/R_0\right)^2+(s/R_0)^2}} ~\frac{ds}{R_0},
	\end{eqnarray}
	 we find
\begin{eqnarray}
  \frac{z(\tau)}{h} & \approx & \ln \left( \frac{2\Dh\sigma\rho(0)}{\tau} \right) - 1.
\end{eqnarray}
In general, $h$ may be different for each atmospheric constituent. If
we assume that the components are uniformly mixed throughout the
atmosphere, then $h$ is independent of composition, and the difference
between two heights $z_1(\tau)$ and $z_2(\tau)$ at two distinct
wavelength bands with different opacities $\sigma_1$ and $\sigma_2$
can be written independently of the optical depth, as
\begin{eqnarray}
  \delta z = z_1 - z_2 & \approx & h \ln \left(\frac{\sigma_1}{\sigma_2}\right).
\end{eqnarray}
If we assume further that the temperature scale height is large
compared to the pressure scale height, then $\Dh \approx \PH$ and $\NS
\approx \ln \left(\sigma_1/\sigma_2\right)$.

We may now judge the plausibility of this interpretation with
reference to the typical opacities and widths of molecular absorption
features. For strong molecular bands and atomic lines, the ratio of
the in-band opacity to the nearby continuum opacity may be as large as
$10^4$ (Brown~2001, Seager \& Sasselov~2000), yielding a maximum
height difference of $\NS \approx 10$ scale heights within the
absorption band. The NICMOS band from $1.1-2.0~\mu$m includes strong
rotation-vibration molecular absorption bands due to H$_2$O, CO and
CH$_4$. An example of a very strong absorption band is a water band
centered at $1.4~\mu$m, spanning approximately $10\%$ of the effective
filter width. If we assume that this is the dominant spectral feature
in this band, then the result of $\NS\approx 5$ across the entire
bandpass translates into $\NS \approx 50$ within the bandpass of the
absorption feature. This is larger than the criterion $\NS \approx 10$
mentioned above.

Therefore this interpretation seems to require significantly more
opaque or broader-band absorption features than are seen in the
models. In one sense the result of the order-of-magntiude calculation
is discouraging, as it may make it seem more likely that the
discrepancy in depths is due to systematic errors. On the other hand,
if the noise were well-understood and the discrepancy could be
confidently proclaimed, then it would be the sign of new and
interesting atmospheric physics that is not described in the standard
models. Some priorities for progress on this issue include an
examination of the wavelength-dependence of the transit across the
NICMOS band, and the observation of the system with other NICMOS
grisms, which are reputed to be more stable than the G141 grism used
here.

\acknowledgements We thank G.~Torres for helpful discussions
concerning the determination of stellar parameters. This work was
supported by NASA grant HST-GO-11165 from the Space Telescope Science
Institute, which is operated by the Association of Universities for
Research in Astronomy, Incorporated, under NASA contract NAS5-26555.

\newcommand{\degdot}{\stackrel{\circ}{.}}

\begin{table}[htbp]
   \centering
     \begin{tabular}{lllll}
     	\hline \hline
     	 & &\multicolumn{3}{c}{Simultaneous Fit} \\
	 \cline{3-5}
	Parameter 						&  NICMOS only 						& NICMOS 							& {\em Spitzer}\, 8~$\mu$m 					& $(b+y)/2$ \\
	\hline \\
	Wavelength range [$\mu$m] 			& 1.1--2.0 								& 1.1--2.0 								& 6.5--9.5									& 0.45--0.55 \\ 
	Cadence [secs]\tablenotemark{a} 			& 7.2 								& 7.2 								&  4.1 									& 8.6 \\ 
	Normalized error\tablenotemark{b} 			& 0.00044 							& 0.00044 							& 0.0026 									& 0.0020 \\\\
   	$R_p/R_\star$ 						& $0.05416^{+0.00091}_{-0.00070}$ 		& $0.05430^{+0.00085}_{-0.00078}$		&$0.05188^{+0.00084}_{-0.00086}$			&$0.05070^{+0.00058}_{-0.00088}$ \\
	$(R_p/R_\star)^2\times100$ 			& $0.2933^{+0.0099}_{-0.0076}$ 			& $0.2949^{+0.0092}_{-0.0085}$ 			& $0.2692^{+0.0087}_{-0.0089}$ 				& $0.2570^{+0.0059}_{-0.0089}$ \\
	$i$ $[{\rm deg}]$ 					& $84.55^{+0.35}_{-0.81}$ 				& $84.50^{+0.60}_{-0.52}$  				&  										&  \\
	$a/R_\star$ 						& $6.01^{+0.17}_{-0.23}$ 					& {$5.99^{+0.21}_{-0.18}$} 				&  										& \\
	$b \equiv a \cos i/R_\star$ 			& $0.571^{+0.044}_{-0.038}$ 				& $0.574^{+0.035}_{-0.045}$ 				&  										&  \\ 
	Ingress Duration $[{\rm hr}]$\tablenotemark{c} 	& $0.241^{+0.012}_{-0.013}$ 				& $0.243^{+0.012}_{-0.013}$ 				& $0.232^{+0.011}_{-0.012}$					& $0.227^{+0.011}_{-0.012}$ \\
	Transit Duration $[{\rm hr}]$\tablenotemark{d}	& $3.24^{+0.14}_{-0.15}$ 					& $3.25^{+0.14}_{-0.15}$ 					& $3.24^{+0.14}_{-0.15}$ 						& $3.23^{+0.14}_{-0.15}$\\ \\
	
	$M_\star$ $[M_\odot]$ 				& $1.345^{+0.020}_{-0.020}$ 				& $1.34^{+0.020}_{-0.020}$ 				&  										&  \\
	$R_\star$ $[R_\odot]$ 				& $1.541^{+0.046}_{-0.042}$ 				& $1.534^{+0.049}_{-0.047}$ 				&  										&  \\
	$\rho_\star$  [g~cm$^{-3}$] 			& $0.497^{+0.042}_{-0.057}$ 				& $0.492^{+0.052}_{-0.044}$ 				&  										&  \\
	$\log g_\star$ [cgs] 					& $4.189^{+0.020}_{-0.021}$ 				& $ 4.192^{+0.022}_{-0.022}$ 				&  										&  \\
	Distance [pc] 						&  $83.0^{+2.8}_{-2.7}$ 					&  $82.6^{+2.8}_{-2.8}$ 					&  										&  \\
	Stellar Age [Gyr] 					& $2.6^{+0.3}_{-0.2}$ 					&  $2.6^{+0.2}_{-0.2}$ 					&  										&  \\
	$L_\star$ $[L_\odot]$ 				& $3.03^{+0.20}_{-0.18}$ 					&  $3.00^{+0.20}_{-0.19}$ 				&  										&  \\ \\
	
	$M_p$  $[M_{\rm Jup}]$\tablenotemark{e} & $0.368^{+0.013}_{-0.014}$ 				& $0.366^{+0.014}_{-0.013}$ 				&  										&  \\
	$R_p$  $[R_{\rm Jup}] $ 				& $0.813^{+0.027}_{-0.025}$ 				& $0.811^{+0.029}_{-0.027}$ 				& $0.775^{+0.028}_{-0.027}$ 					& $0.757^{+0.025}_{-0.027} $ \\
	$\rho_p$  [g~cm$^{-3}$] 				& $0.85^{+0.10}_{-0.09}$ 					& $0.85^{+0.10}_{-0.09}$ 					& $ 0.98^{+0.11}_{-0.11}$ 					& $1.05^{+0.11}_{-0.12}$ \\ 
	$\log g_p$ $[{\rm cgs}]$\tablenotemark{e} 		& $3.132^{+0.029}_{-0.035}$ 				& $3.127^{+0.033}_{-0.029}$ 				& $3.166^{+0.034}_{-0.030}$ 					& $3.186^{+0.032}_{-0.030}$\\
	\\\hline
   \end{tabular}
   	\caption{ System Parameters of HD~149026.}
     \label{tab:MCMCout}
   \tablenotetext{a}{Defined as the median time interval between data points in the composite (phase-folded) light curve.}
   \tablenotetext{b}{Defined as $\sigma_F/F$ where $\sigma_F$ is the rms residual between the data and best-fitting model, and $F$ is the out-of-transit flux. }
   \tablenotetext{c}{Defined as the time between first and second contacts, or between third and fourth contacts. (In our model these durations must be equal.)}
   \tablenotetext{d}{Defined as the time between the first and fourth contacts.}  
   \tablenotetext{e}{Using $K = 43.3\pm1.2$ m~s$^{-1}$, from Sato et al.~(2005).  }

\end{table}

\begin{table}[htbp]
   \centering
     \begin{tabular}{lll}
     	\hline \hline
	Epoch & Mid-transit time [HJD] & Error \\ \hline
	$0$ & $2454456.78751$ & $0.00030$ \\
	$1$ & $2454459.66379$ & $0.00023$ \\
	$17$ & $2454505.67688$ & $0.00034$ \\
	$31$ & $2454545.94133$ & $0.00047$\\ \hline
  \end{tabular}
   \caption{Mid-transit times, based on the NICMOS data.}
   \label{tab:times}
\end{table}

\end{document}